\pdfoutput=1 
\documentclass{JINST}
\usepackage{amsmath}

\title{Performance study of the Jalousie detector baseline design for the ESS thermal powder diffractometer HEIMDAL through GEANT4 simulations}

\author{I. Stefanescu$^a$\thanks{Corresponding author.},   M. Christensen$^b$, R. Hall-Wilton$^{a}$, S. Holm-Dahlin$^{c}$, K. Iversen$^b$, M. Klein$^d$, D. Mannix$^a$, J. Schefer$^e$, C. J. Schmidt$^{d,f}$, W. Schweika$^{a,g}$, U. Stuhr$^e$\\
\llap{$^a$}European Spallation Source ESS ERIC, P.O. Box 176, SE-22484, Lund, Sweden\\
\llap{$^b$}Center for Materials Crystallography iNANO$\&$Department of Chemistry, Aarhus University,
DK-8000 Aarhus, Denmark\\
\llap{$^c$}ISIS Department, Science and Technology Facilities Council, Rutherford Appleton Laboratory,
Harwell Oxford, Oxfordshire, OX11 0QX, U.K.\\
\llap{$^d$}CDT CASCADE Detector Technologies GmbH, Hans-Bunte-Str. 8-10, 69123 Heidelberg, Germany\\
\llap{$^e$}Laboratory for Neutron Scattering and Imaging, Paul Scherrer Institute, 5232 Villigen, Switzerland \\
\llap{$^f$}GSI Detector Laboratory, Planckstrasse 1, 64291 Darmstadt, Germany\\
\llap{$^g$}Forschungszentrum J\"ulich GmbH, 52425 J\"ulich, Germany\\
E-mail: \email{irina.stefanescu@esss.se}}

\abstract{HEIMDAL is a thermal powder diffractometer designed to operate at the European Spallation Source, world's most intense neutron source. The detailed design of the instrument, which is expected to enter user operation in 2024/2025, assumes that the neutrons scattered by the powder under investigation will be collected with hundreds of large-area Multi-Wire Proportional Counters employing a $^{10}$B$_4$C-solid converter. The gas counters will consist of large active volumes and tapered trapezoidal shapes that allow for close packing into a cylindrical shell with high solid angle coverage. The whole detector will operate in an air environment within the shielding cave and provide signals with sensitivity for locating detection in three dimensions. This paper presents the results of a GEANT4 study of the baseline design for the HEIMDAL powder diffraction detector.  The detector model was used to study the spatial resolution, which in the horizontal scattering plane must be below 3 mm in order to enable  high-resolution diffraction studies. The contribution of the detector to the resolving power of the instrument, one of the key figures-of-merit for powder diffractometers,  was also investigated.  Most of the simulation results reported in this work cannot be validated against a sufficiently similar physical reference until the first segment or module are constructed and tested with a neutron beam. However, these results can help to identify possible ways of optimising the detector design and provide the first glimpse into the expected performance of this technological approach.}


\begin{document}

\section{Introduction}\label{sec:intro}

Neutron scattering continues to make important contributions to society via the development of new materials for life sciences and industrial applications. Progress in this field requires the existence of strong neutron sources and advanced instrumentation. European Spallation Source (ESS), currently under construction in Lund, Sweden \cite{ess}, and the planned upgrades of the existing ISIS \cite{isis,Jones1,Jones2} and SNS \cite{sns_tdr} spallation sources will give the scientific community the opportunity to access high-intensity neutron beams, making it possible to find solutions to challenging problems in modern physics and chemistry that go far beyond the traditional research programmes. 

ESS will be a low repetition rate, long-pulsed source with a pulse length of 2.86 ms at a repetition rate of 14 Hz and capable of delivering both cold and thermal neutron beams  \cite{tdr,garoby}. When finalised, the source is expected to attain a peak neutron brightness of $\sim$ 8$\cdot$10$^{14}$ n$\cdot$cm$^{-2}\cdot$s$^{-1}\cdot$sr$^{-1}$$\cdot$\AA$^{-1}$ for both the cold and thermal spectra ~\cite{tdr,garoby}. At completion, a total of 42 beam ports will be made available for research, but there will be a staged approach to selection and construction. The first 15 instruments will be coupled to the ESS neutron source during the construction and initial operation phases (2014 - 2025), with the goal to start the user operation in 2023 \cite{act_rep,nid}. The selection process for an additional 7 instruments is planned to start soon \cite{gap}.   All instruments will be built collaboratively by researchers from institutions across Europe and in-house scientists. 

The  ESS  instruments  are expected to gain 10-100  times  over  current  performance,   enabling  neutron  methods  to  study  real-world  samples  under  real-world  conditions in real-time \cite{nid}. This is expected to have a radical effect on the way many studies are carried out, and requires performant instrumentation. For example, neutron guides to transport the intense beam efficiently and detectors able to deal with the high event rate and capture the position of the scattered neutrons with better accuracy than what is possible with the existing technologies \cite{vertex}. 

Neutron diffraction is an important tool for material research and life sciences. Powder- and single-crystal diffractometers are among the beamlines  selected to enter the user operation in the early days of every new neutron scattering facility.  At ESS, the bi-spectral powder diffractometer for DREAM ({\bf D}iffraction {\bf R}esolved by {\bf E}nergy and {\bf A}ngle {\bf M}easurement) \cite{dream,dream1}, the thermal powder diffractometer HEIMDAL (named after the Old Norse god {\it Heimdall} \cite{wiki}) \cite{heimdal,Holm} and the single-crystal diffractometer MAGIC ({\bf MAG}net{\bf IC} single crystal diffractometer) \cite{magic} were selected by the Scientific Advisory Committee to deliver the most essential capabilities required by the neutron crystallography user community \cite{instr}. These instruments must operate at an internationally competitive level and are designed to complement and extend the existing scientific capability of the neutron powder diffraction methods. The design drivers for the ESS neutron diffraction instruments are the efficient use of the bright ESS neutron source, the flexibility for choosing between high resolution and high flux modes of operation with good resolution, and the access to as high a {\it Q}-range\footnote[1]{Q is the magnitude of the scattering vector (transferred momentum) for elastic scattering i.e., assuming no change of neutron energy in the scattering process. It is defined as $Q=(4\pi/\lambda)\cdot\sin\theta$, where $\lambda$ is the neutron wavelength and $\theta$ is the scattering angle.} as possible. These impose challenging requirements on the detectors, readout electronics and data handling necessitating an improvement of the current technologies. The technical specifications for the detectors that will be deployed at the future ESS diffraction instruments along with the baseline options are documented in Ref. \cite{Ste03}.  At the time of writing, all three neutron diffraction instruments passed the Preliminary Design Review and entered the detailed engineering design phase. Documentation covering the system description, requirements for each of the subsystems, performance expectations, cost estimates and preliminary schedules were made available by the collaborating institutions and are currently under review by the technical teams within ESS. 

The detector scope described in the documents submitted for instrument project review covers all instrumentation required to measure the time and position of the scattered neutrons (i.e., detector hardware and associated readout) and the beam monitors to measure the characteristics of the beam. The tenfold increase in performance required by ESS demands a detector technology with count-rate and position resolution capabilities beyond what is currently possible with pressurised $^3$He-tubes \cite{Ste03}. One of the technologies that appears to fulfil most of the requirements set by the DREAM, MAGIC and HEIMDAL instrument teams is the $^{10}$B-based Jalousie detector \cite{Hen12} developed commercially by CASCADE Detector Technologies (CDT) in Heidelberg, Germany \cite{cdt}.  

In this paper we present the results of the GEANT4 performance study of the baseline design for the Jalousie detector for the HEIMDAL thermal powder diffractometer \cite{heimdal}.  In order to be able to study the spatial resolution of the detector, one of the key technical specifications, the model for the sensitive area was built of small 3D-gas volumes that account for the voxelized readout structure of the real system. This simple method of coding the geometry of the detector ensures easy navigation through the different volumes and fast computing times without affecting the realism of the modelled physical system. The gas-voxel model was validated by comparing the results of the simulation with the experimental measurements to determine the position resolution of one of the early prototypes of the Jalousie segment.  Furthermore, we used the Jalousie detector model  to study of the contribution of the baseline detector design to the HEIMDAL instrumental resolution function, the figure-of-merit that characterises the performance  of a diffraction instrument with respect to its resolving power. This was done by simulating the diffraction pattern of a sample with well-known crystallographic structure and analysing the widths of the recorded Bragg peaks. The simulations were performed  with both the HEIMDAL detector model and the model of a reference detector consisting of a cylindrical array of $^3$He-position sensitive tubes, one of the state-of-the-art technologies in use at the existing neutron diffraction instruments. The comparative study of the detector contribution to the Bragg peak width assesses the performance and viability of the Jalousie detector technology for neutron powder diffraction as well as its relevance and potential for use in a large range of neutron scattering applications. 

\section{Neutron powder diffraction}\label{sec:npd}

Neutron powder diffraction is a powerful tool to study materials that are composed of many different crystallites. The powder can be represented by many different crystal lattices planes separated by the distances $d_i$ and at various orientations. The interaction of a beam of neutrons with the powder produces constructive interference and intensity cones when Bragg's law $\lambda_i=2d_i\sin\theta$ is satisfied (see footnote 1 for the definition of the parameters $\lambda$ and $\theta$). These cones will intersect a detector with a flat surface as circles, known as Debye-Scherrer rings of diffraction \cite{pdf}. The analysis of the detected scattered intensity gives information about the structure, strains, defects and other aspects of the crystal under a range on experimental conditions.  

Data collected in powder diffraction measurements is conveniently displayed as 1D-spectra representing counts against  {\it time-of-flight} or $d$-spacing. This leads to discrete peaks called Bragg reflections corresponding to the different $d_i$ lattice planes in the material under investigation. The set of $d$-spacings and intensities obtained from a diffraction pattern for a random powder sample of a given material is unique. The Bragg peaks can be fitted to the sum of reflected intensities of an assumed crystal structure, broadened by the instrumental and sample effects of individual reflections plus the underlying background. Such analysis procedure is known as \emph{Rietveld profile refinement} \cite{rietveld}. The fitted position and width of the peaks are used to extract the lattice size and provide information on the crystallite structure and strain effects respectively, while the fitted intensities provide information about the atomic fractional coordinates, atomic occupation factors and thermal vibration.  

The standard for detection in modern powder diffraction instruments is the employment of large, contiguous detectors with cylindrical geometry around the sample \cite{cussen}, which makes it possible to collect simultaneously a portion of all diffraction cones. Additional detectors at low (forward) and large (back-scattering) angles are also used for texture determination for non-axis symmetric features or strain/dislocation density and magnetic studies,   respectively. 

Almost all powder diffraction instruments operational today use large banks or panels of pressurized $^3$He-filled tubes or scintillator-based position-sensitive detectors for recording the scattered neutrons. A short overview of the state-of-the-art detector technologies employed at existing neutron diffraction instruments is presented in Ref. \cite{Ste03}. The detector banks or panels cover different $\theta$-regions around the sample under investigation and record different ranges of $d$-spacing. Good resolution is desired as it minimises the overlapping of adjacent diffraction peaks and allows for the observation of well defined individual features, which is essential for a precise solving or refinement of the crystalline structures.  The most used range of wavelengths is between 0.5 and 4 \AA, which coincides with the region where most of the samples show a high density of Bragg reflexions. Therefore, the detector must be designed for the optimal detection of neutrons with wavelengths matching the range of interest.      



\subsection{The thermal powder diffractometer HEIMDAL}\label{subsec:instr}

HEIMDAL will be a thermal powder diffractometer for use for chemistry, crystallography and materials science studies \cite{nid,heimdal}. It will be built by a collaboration between the Aarhus University in Denmark \cite{au}, Institute for Energy Technology (IFE) in Norway \cite{ife} and Paul Scherrer Institute (PSI) in Switzerland \cite{psi}. The instrument concept is described in detail in Ref. \cite{Holm}.  HEIMDAL will have a total length of 157 m and it will cover a $Q$-range from 0.5 \AA$^{-1}$ up to 25 \AA$^{-1}$. The instrument will be designed from its onset to support a small-angle neutron scattering and a neutron imaging station. These upgrades will make HEIMDAL one of its kind as it will provide capabilities to perform structural studies over multiple length scales.   

\begin{figure*}[ht]
\centering
\includegraphics[scale=0.6]{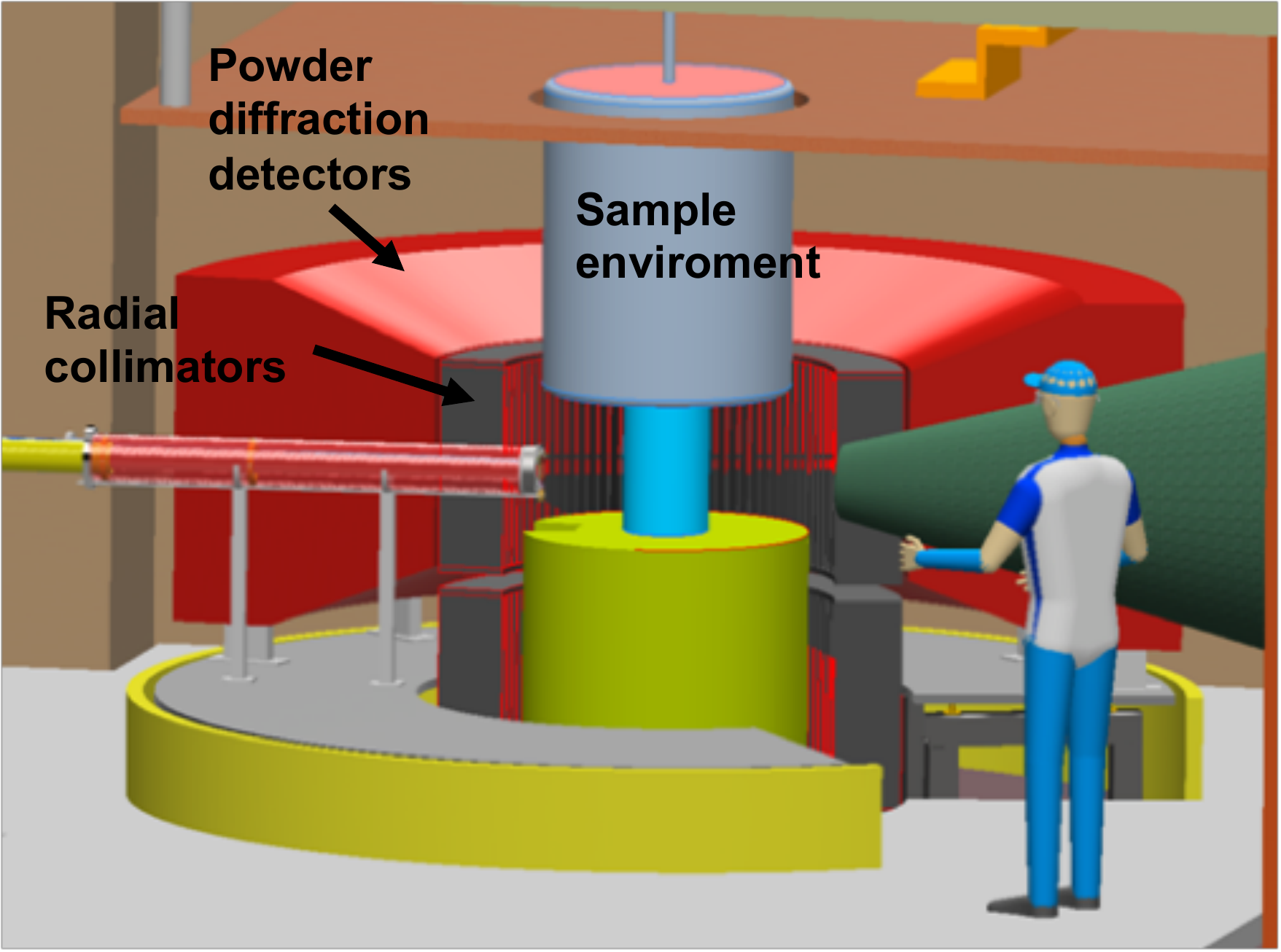}
\caption{Conceptual view of the HEIMDAL sample area. The sample environment, collimators and powder diffraction detectors are marked on the figure. }  
\label{fig:f0}
\end{figure*}

The goal of the HEIMDAL design team is to deliver Day-1 of operation a powder diffractometer that has a high degree of operational flexibility in matching the pulse length with the requirements of the material under investigation. This will be achieved by using a combination of pulse shaping choppers that will select a portion of the long initial neutron pulse ($\sim$3 ms), which will be transported to the powder sample with the help of performant beam optics. HEIMDAL will operate with a relatively narrow wavelength band  ($\Delta\lambda$ = 1.74 \AA) and a mean wavelength $\lambda$ = 1.5 \AA, which is close to the peak brightness of the ESS thermal moderator \cite{Zan}.  The length of the pulse selected by the choppers affects the available flux on the sample. By trading part of the flux with narrower pulses and assuming small contributions from the moderator and beamline optics, high quality powder diffraction data with $\Delta d/d$ down to 9$\cdot$10$^{-4}$ can be collected in the detector located around 90$^\circ$ \cite{Holm}. However, this requires that the position of the interaction is determined with a precision of 3 mm or less, which is beyond the capability of current detector technologies  \cite{Ste03}.  
 
A detailed discussion of the requirements for the detectors for HEIMDAL is presented in Ref. \cite{Ste03}.  In that paper, it was suggested that scintillator-based detectors is the favored option for this instrument. Unfortunately, the level of R\&D involved in order to reach the required readiness level on time was deemed too high and therefore, not possible in view of the limited resources available. The combination of performance goals, resources and schedule have led the instrument team to consider the $^{10}$B-based detector technology called Jalousie \cite{Hen12}.  This detector technology was developed for use by the reactor-based powder diffractometer POWTEX \cite{powtex} that will operate at the FRM2 reactor in Munich, Germany \cite{frm2}. The requirements of the neutron source, energy range and resolution, specifics of the desired sample environments, scientific goals,  space and logistics constraints and not least, the allocated budget, led to slight differences in the technical specifications for the powder diffraction detectors for these instruments. A discussion of the differences between the various Jalousie designs is beyond the scope of this paper and it will addressed in a future publication.

\subsection{The Jalousie detector}\label{sec:jal}

The Jalousie detector concept was developed to address the $^3$He crisis caused around year 2009 by the large demand for use in various applications corroborated with the decline in the production of the gas \cite{karl}. In view of the projected need for the new instruments at the future European Spallation Source and planned upgrades of the existing neutron scattering facilities, several research labs and private companies worldwide devoted resources to develop alternatives for large-area detectors \cite{powgen,gem}. The potential of the boron thin film gaseous detector, one of the oldest technology used to detect thermal and cold neutrons, was immediately recognised and now, almost a decade later, gas detectors with solid $^{10}$B$_4$C converter layers are being considered for deployment in over half of the initial instruments at ESS \cite{mg,mb}. 

The Jalousie detector design also exploits the  traditional proportional wire counter and the $^{10}$B$_4$C-solid converter.  The counters are arranged in cylindrical geometry around the sample such that the detecting surface is curved around the Debye-Scherrer rings of diffraction and tilted with respect to the diffracted neutron path in order to fulfil the requirement for high detection efficiency \cite{Ste03}.  The neutron detection is achieved in the Jalousie detector via the capture reaction $^{10}$B(n,$\alpha$)$^7$Li, which has a cross-section of 3840 barns for neutrons with a wavelength of  $\lambda$=1.8~\AA. The two reaction products are emitted back-to-back following the conservation of momentum. The charged particle released in the counting gas creates electron-ion pairs. The electrons are accelerated toward the wires in predetermined paths given by the electric field created by applied a high voltage to the anode wires. Near these, the ionisation electrons undergo avalanche multiplication resulting in a large number of secondary electron-ion pairs that induce an electrical signal on the anode wire and a reciprocal, inverted signal on the cathode, which is kept at ground potential. The detector is designed to operate in proportional mode in continuous flow of Ar/CO$_2$ gas (90-10). 

\begin{figure*}[ht]
\centering
\includegraphics[scale=0.8]{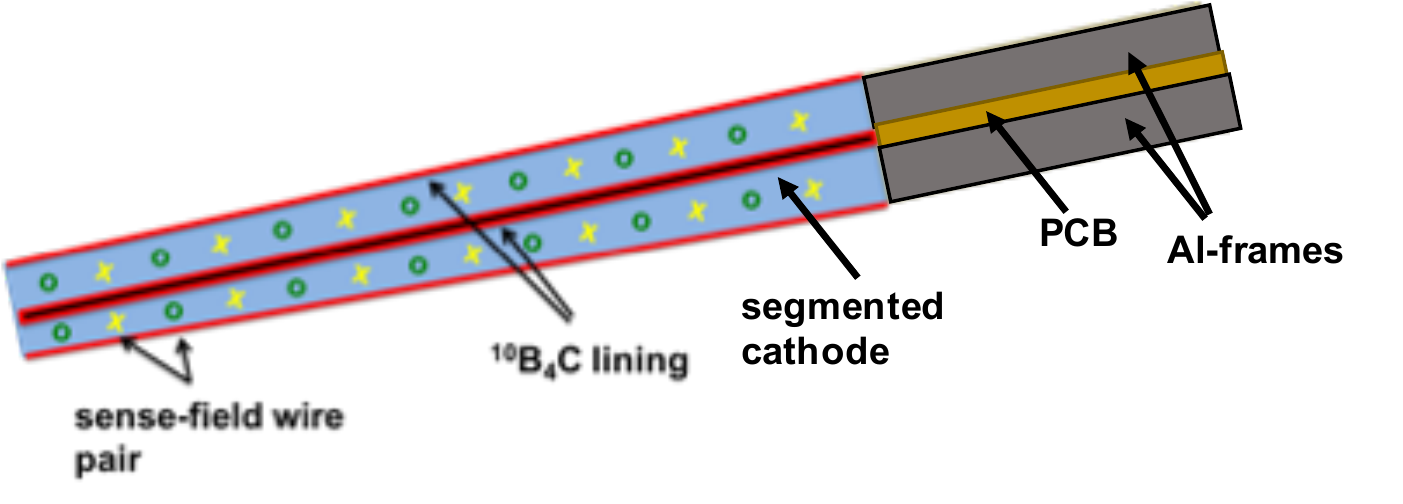}\\
\includegraphics[scale=0.9]{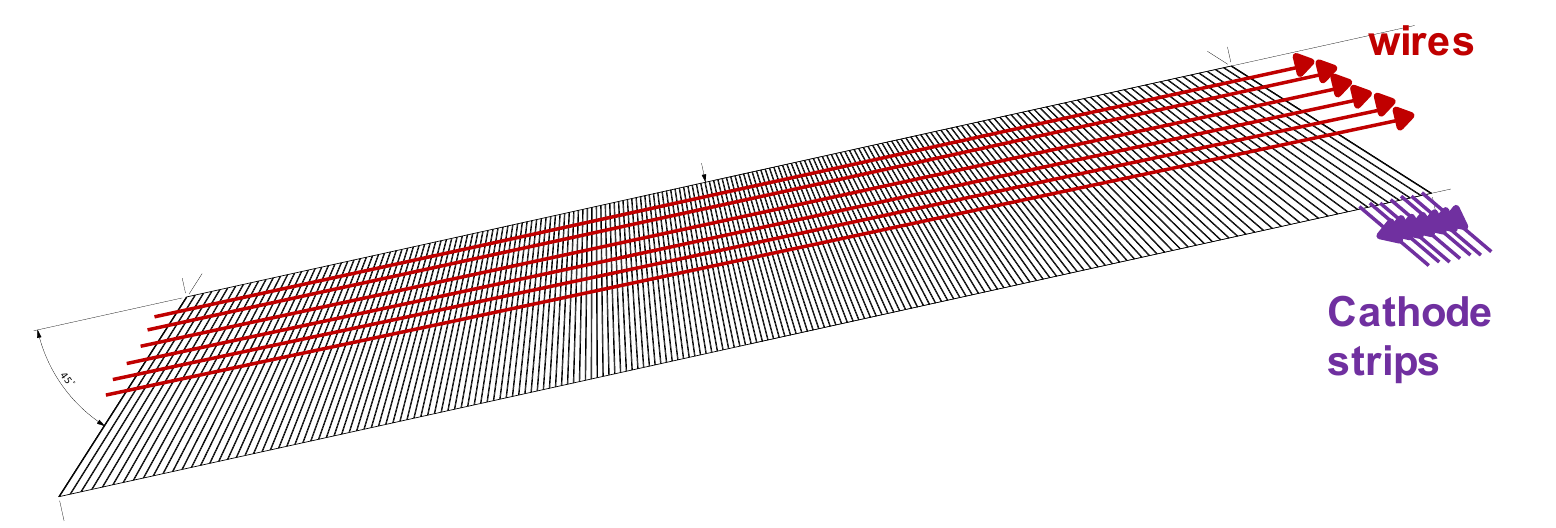}
\caption{Schematic view of a generic Jalousie segment. Top: Cross-sectional view showing the segmented cathode in the middle, the $^{10}$B$_4$C lining (red), the alternating pattern of the sense (green open circles) and field wires (yellow cross). The wires are shown end-on. The tapered shape of the segment ensures a high packing factor in the cylindrical frame. The PCB (printed circuit board) shown in orange hosts the electrical connections from the cathode strips and it is sandwiched between two Aluminium support frames, which are part of the larger frame assembly that ensures the mechanical stability of the whole detector segment. Bottom: drawing showing the segmented cathode and the sense wires running perpendicular to the strips. The arrows indicate individual readout.}  
\label{fig:f1}
\end{figure*}

The basic Jalousie detector unit is a thin trapezoidal gas counter called segment \cite{Hen12}. The segment housing is manufactured from an Aluminium sheet with a thickness of 300 $\mu$m.  Each segment will host two  independent wire counters  separated by a trapezoidal-shaped cathode, see the cartoon shown in Fig. \ref{fig:f1}. The wire grid consists of an alternating pattern of sense and field wires. The field wires are at ground-potential and their role is to optimally shape the electrical drift field inside the counter and remove the crosstalk between transient signals of charge collected on the anode wires. The cathode is segmented into strips that are sensitive to the charge induced by the ionisation electrons. The wires and the cathode strips are positioned perpendicular to each other. The size and the number of strips as well as the number of sense-field wire pairs per counter is different for the different variants of the Jalousie detector and it depends on the requirements for spatial resolution defined by the instrument teams. However, the wires and cathode strips will be readout independently in all of them, as sketched in the bottom panel of Fig. \ref{fig:f1}. This ensures high-rate capability for neutron detection, which is one of the key requirements for all detectors that will be deployed at ESS \cite{vertex}. The board hosting the electrical connections for the cathode strips and the readout electronics are located at the rear of the segment. 

The geometry of the gas counter and electric field as well as the employed readout scheme give rise to an active area consisting of a grid of 3D sensitive cells (voxels). The volume of each voxel is given by the product between the wire pitch, strip size and the gas gap, where the latter is determined by the distance between the detector housing and the segmented cathode. This makes the Jalousie detector a highly efficient 3D position-sensitive detector that is able to provide information on the depth of the interaction point for each detected neutron. 

The cylindrical geometry specific to detection systems used by diffraction instruments will be obtained by mounting  hundreds of segments side-by-side in a fan-like structure around the sample under investigation. The segments are designed with a tapered shape that allows   close packing into a cylinder around the sample.  For the POWTEX and DREAM mantel detectors, the design is such that the cylinder will have its axis in the beam direction, providing hermetic and uniform $\phi$-coverage around the sample under investigation and 90$^\circ$ acceptance in 2$\theta$ \cite{dream,powtex}.  In the HEIMDAL instrument, the detector cylinder will have the symmetry axis perpendicular to the beam axis and cover an angle $\phi$=$\pm$22$^\circ$ in the vertical plane, see next subsection. 

To facilitate the installation work, maintenance and testing, the Jalousie detector is designed to be modular. A number of detector segments are assembled in a common mounting unit. The Jalousie  module for HEIMDAL will feature a number of six geometrically identical segments. Assembling the necessary detection volume out of closely spaced segments and modules has many practical advantages, allowing their production to be standardized with clear benefits on quality and costs. The modular design also allows for incremental ramp-up of detector coverage and operation, since even a fraction of the final solid-angle makes it possible to test the performance of the beamline and enables early results. 

The detection efficiency is a design parameter of prime importance for all applications. State-of-the-art neutron detector technologies employed at existing neutron scattering facilities  are able to detect the diffracted signal with an efficiency above 50$\%$ at 1.8~\AA. In order to achieve the same or better performance, the Jalousie segments are mounted in the frame such that the angle between the incoming neutron and the Boron layer is 10$^\circ$. Turning the Boron layer at an angle is a method that can be used to increase its efficiency because it maximises the path length of the incident neutrons through it, whilst keeping it thin enough to allow escape of the conversion products. However, a significant increase in the efficiency of one layer occurs only for incidence angles less than 30$^\circ$ (normal incidence is at 90$^\circ$) \cite{Hen12}. Therefore, the tilt angle chosen by the detector designer is the best compromise between the gain in efficiency and mechanical design, which becomes more challenging as the solid angle covered by the Boron layer decreases with decreasing the incidence angle. 

In the generic Jalousie detector a Boron layer with a thickness between 1 and 1.2 $\mu$m is applied on the inner faces of the segment housing and on both sides of the segmented cathode. This results in four Boron layers per detector segment, which limit the efficiency for detecting thermal neutrons (1.8 \AA) to values below 50$\%$. For this specific detector design the requirement for higher efficiency can be fulfilled by choosing the depth of the segment  such that each incoming neutron must pass at least 8 Boron layers (i.e., two detector segments)  \cite{Hen12}. 

The first proof-of-concept studies were made for the baseline detector design for the POWTEX instrument and date back to 2011-2012 \cite{ModzelT,Modzel}. A so-called ``prototype-0'' was built to evaluate the performance of a single segment. After the selection of the Jalousie technology by the technical team in charge of building the new powder diffraction instrument POWTEX \cite{powtex}, the design of the detector components has been refined in view of setting up an assembly line  production at the CDT premises, using the experience gained with the prototype-0.  The assembly line production concept was validated by the construction and testing of a second prototype segment, which used final or close to final series elements for its construction \cite{Hen12}.  
At the time or writing this paper, CDT is in the process of finalising the manufacturing the detector modules for the POWTEX instrument and close to finalising the design details for the segments for the DREAM instrument. 

\subsection{The Jalousie detector for HEIMDAL}\label{subsec:jal_h}

As both the HEIMDAL instrument and the detector are still in design phase, the dimensions and geometry of  the Jalousie mantel detector discussed here are preliminary and subject to refinement.  The detector specifications used in this work represent the baseline design submitted by the instrument team to the latest instrument review held at ESS. It represents the best guess that rests heavily on the experience and lessons learnt during building and testing the Jalousie detector modules for POWTEX.  Specific details about how the final detector will look like in reality and how challenges associated with the engineering design will be mitigated are currently being discussed and reviewed.  

The Jalousie segments for the HEIMDAL powder diffraction detector will lie on an arc with a radius of 80 cm and have a height of 640 mm at the entrance window. These dimensions were chosen to match the opening of the magnet and pressure cell in the sample environment \cite{Holm}, and at the same time cover $\phi$=$\pm$22$^\circ$ with no discontinuity along the azimuthal angle.  At completion, the diffraction detectors will cover the angular range from 10$^\circ$ to 170$^\circ$ and -60$^\circ$ to -170$^\circ$, but only a reduced angular range will be available in the Day-1 of operation of the instrument. 

\begin{figure*}[ht]
\centering
\includegraphics[scale=0.65]{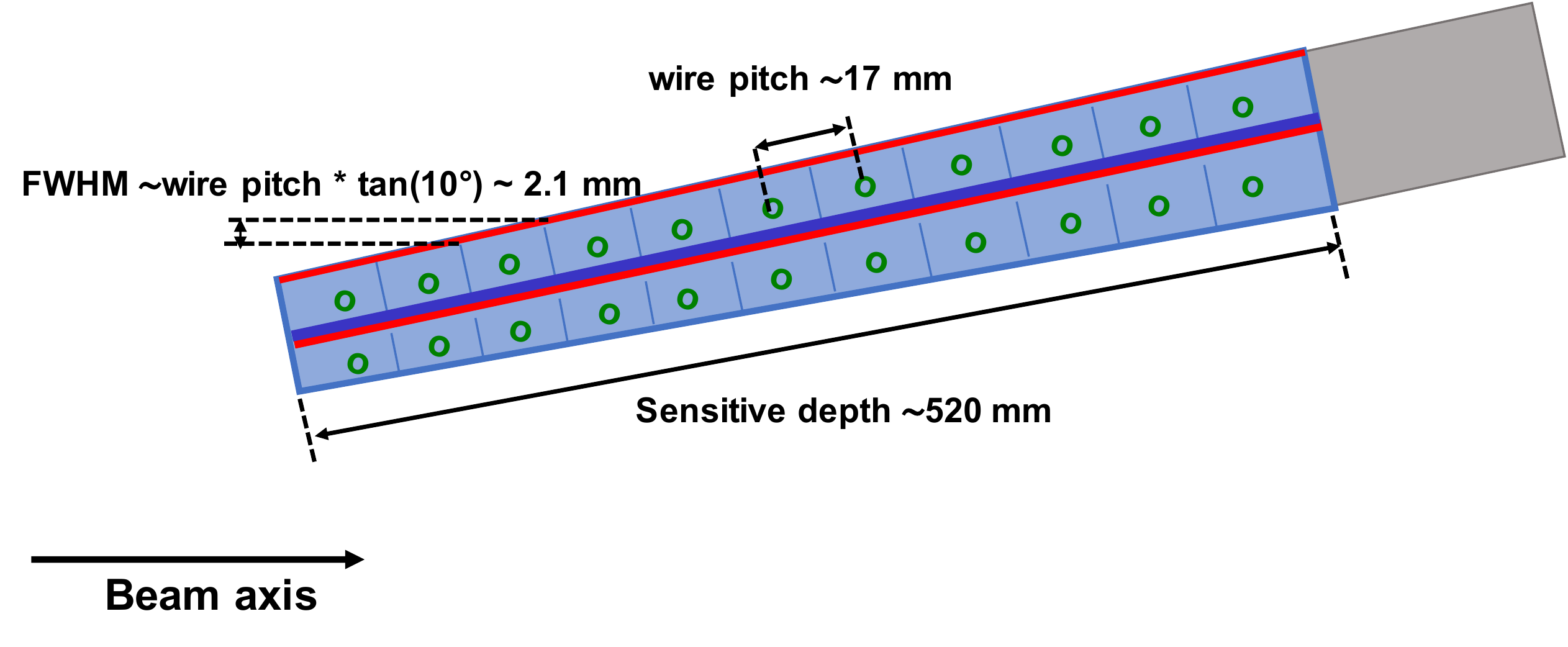}\\
\caption{Cross-sectional view of the HEIMDAL Jalousie segment, showing the Boron layer  in red, the Aluminium substrate and the counting gas in dark and light blue, respectively. The sense wires are represented with open green circles and the field wires with black thin lines. The segment is mounted in the detector frame at an inclination angle of 10$^\circ$ with respect to the beam direction. The horizontal position sensitivity obtained in this arrangement is determined by the projection of the coated surface onto the entrance window. }
\label{fig:f2}
\end{figure*}

In the HEIMDAL version of the Jalousie detector, the Boron coating will be applied on only one side of the wire grid, that is the inner side of the detector housing in the upper counter and on the cathode in the bottom counter or vice-versa, as shown in Fig \ref{fig:f2}. When the segment is mounted in the detector frame at 10$^\circ$ with respect with the beam axis, the projection of the coated side on the entrance window of the detector is proportional to the product  between the  wire pitch and  $\tan(10^\circ)$. This gives rise to a horizontal spatial resolution of $\sim$2.1 mm (FWHM) and an effective $\Delta$2$\theta$-binning of 0.15$^\circ$  resulting from a wire pitch of around 17 mm in depth of the detector segment, in agreement with the specifications by the HEIMDAL instrument team \cite{Holm}. Furthermore, in order to achieve the goal for a detection efficiency greater than 50$\%$ at 1~\AA, the incoming neutron must be able to cross at least eight Boron layers, each having a thickness of 1 $\mu$m and tilted by 10$^\circ$ with respect to the direction of the incoming beam. CDT has developed an analytical method to estimate the depth of the Jalousie-type detectors and the number of Boron layers that need to be crossed by the incoming neutron in order to satisfy the requirement for detection efficiency. This method, the results of the analytical calculations to determine the detection efficiency as a function of realisation depth and the comparison with the  GEANT4 simulations by using the detector model described in this work will be presented in detail in a forthcoming publication \cite{eff_paper}. Analytical calculations indicate that a minimum sensitive depth of 520 mm is needed in order for the Jalousie baseline design  for HEIMDAL to comply with the specifications for the detection efficiency, in good agreement with the results of the GEANT4 simulations \cite{eff_paper}. 

In the baseline segment design the charge generated by the neutrons captured at different depths in the segment will be collected by 32 sense-field wire pairs positioned perpendicular to the beam axis and read out independently.  The cathode will be segmented into 64 strips that will be sensitive to the charge induced by the ionisation electrons. The segmentation of the cathode will be done such that the strips point concentrically to the sample and deliver a constant resolution of $\Delta\phi$=0.68$^\circ$ in the vertical plane. This leads to cathode strips that vary from 9.6 mm at a radius of 80 cm to 17.9 mm at a radius of 1350 mm for $\Delta\phi$ = 0$^\circ$ and from 11.1 mm to 20.8 mm for $\Delta\phi$ = $\pm$22$^\circ$, respectively. This variation with the distance from the sample and the tapered shape of the detector segments in order to achieve close packing and cylindrical geometry lead to detector voxels with widths varying from 7.8 mm at the side close to the sample to 14 mm at the back of the segment. These geometrical aspects could introduce small non-uniformities in the gain of the MWPC along the strip length.  In order to correct for these effects, CDT plans to use 2-4 power supply channels to apply the high voltage on the anode wires, which will be individually adjusted such that the electric field is as uniform as possible over the whole depth of the detector segment.

\section{GEANT4 Simulations}\label{g4_section}

The GEANT4 toolkit is the high-energy physics standard for simulating complex experimental setups and detector responses \cite{g4,g4a,g4b}. GEANT4  is indispensable to the scientific community that designs and builds large hadron colliders and the associated instrumentation. The quest for new neutron detector technologies prompted at the beginning of this decade by the $^3$He crisis  led to a rapid increase of the number of applications of the GEANT4 model for the study of the performance of thermal neutron detectors, especially those based on the $^{10}$B/$^{10}$B$_4$C solid converters \cite{Ste01,Ende,Tsi,Esz}. The increased complexity of the detector technologies under development for the next generation neutron sources such as European Spallation Source makes GEANT4 an indispensable tool that not only provides a quantitative estimate of their capabilities, but it also helps with the optimisation of the design by comparing different technology options or geometric layouts. Simulations of the neutron detectors that will be deployed at ESS also give important feedback to the performance questions raised by the planned staging scenarios due to budget and schedule constrains.

The simulation of the Jalousie detector for HEIMDAL presented in this work made use of Geant4.10.2 version. The physics processes relevant to the transport and interaction of the neutrons with energies below 20 MeV with the matter, as well as the electromagnetic and stopping physics relevant to the secondary charged particles, were accounted for by the reference list QGSP-BIC-HP \cite{g4f}. This is one of the standard lists for use in hadron inelastic physics applications. The list is provided by GEANT4 with the source code and it makes use of the Binary cascade, pre-compound and various de-excitation models for hadrons and a high precision model for neutrons with energies below 20 MeV. The geometry of the Jalousie detector segment was built with the help of boolean operations of the simple pre-defined GEANT4 3D-shapes, such as boxes and trapezoids,  and defining the physical and chemical properties of the constituent materials of these basic volumes. All major collections of components, i.e.,  the detector segment consisting of 32$\times$64 voxels and the module of 6 segments,  were implemented as assemblies of individual geometrical parts. These structures were placed in their correct position (imprinted) as objects in the world volume filled with a very low density gas defined as the GEANT4 galactic vacuum \cite{g4}.  

A very important requirement for the implementation of the Jalousie segment  in GEANT4 is that the model is able to describe realistically the spatial resolution of the detector. In a gas counter, the spatial resolution is determined by the geometry of the detector, such as the wire pitch, strip size and drift gap, high-voltage applied on the wires and the physical (pressure, density) and chemical properties (the gas composition) of the counting gas. The configuration with sense wires under high voltage alternating with field wires at ground potential and grounded cathode strips leads to field lines that leave the anode wire and terminate on the closest cathode strip. The field lines starting from neighbouring sense wires do not intersect. This gives rise to a 3D-unit cell (voxel) with the sense wire in its center, a width equal to the field (sense) wire pitch and height given by the width of the counter (i.e., distance from the segment housing to the common cathode). Furthermore, the common operation mode of a gas-based thermal neutron detector such as Jalousie does not require information about the shape of the signal produced by the readout electronics in order for a detected event to be deemed valid neutron event (also known as {\it pulse shape analysis technique} \cite{knoll}). For most of the $^{10}$B-based gas detectors it is sufficient to compare the amplitude of the pulse, which is proportional to the energy deposited in the counting gas by the reaction products, to an user-set threshold in order to select the valid neutron events \cite{mg,mb}. 

\begin{figure*}[ht]
\centering
\includegraphics[scale=0.8]{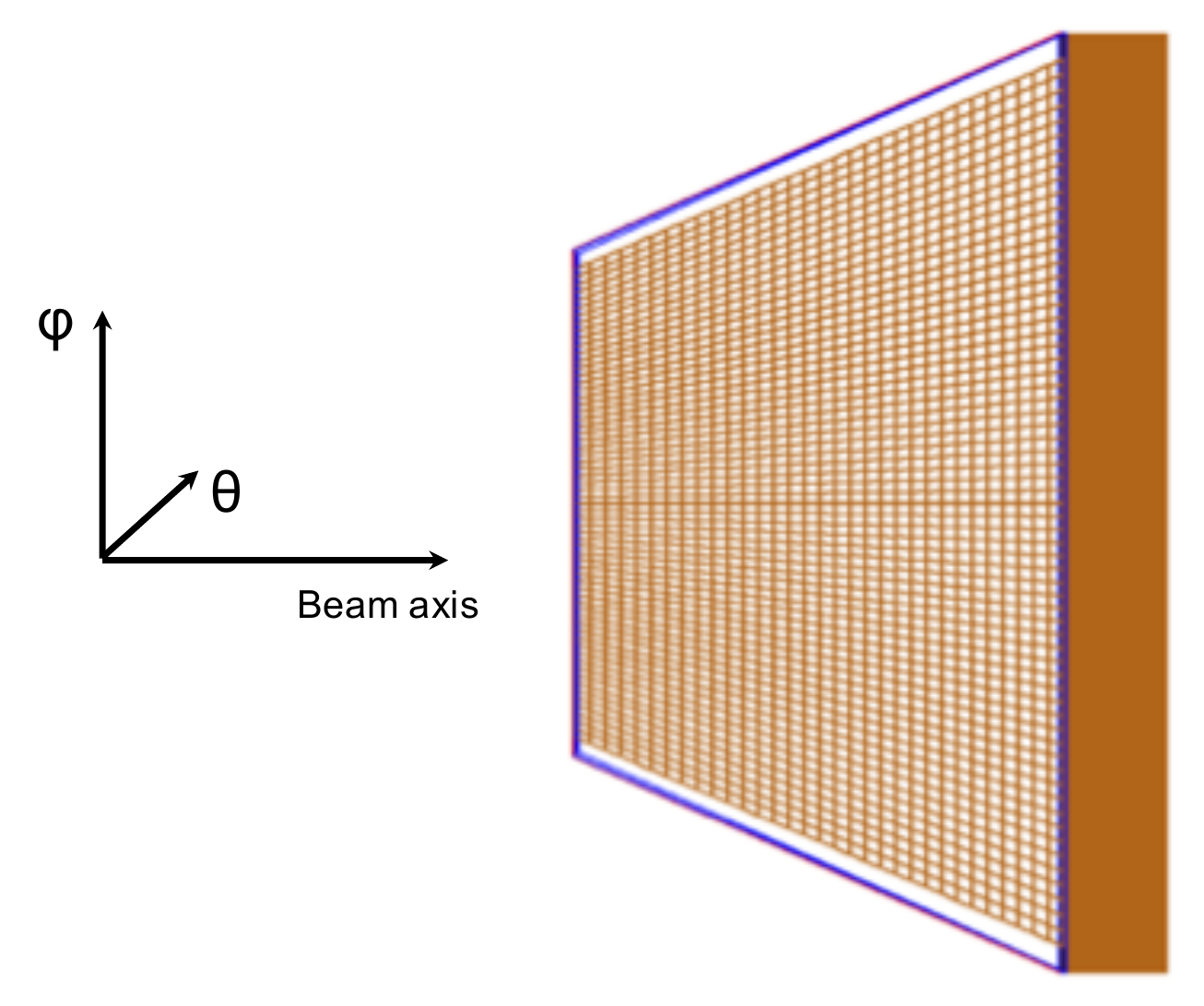}
\caption{ Wireframe representation of  the HEIMDAL detector segment consisting of trapezoidal gas voxels inside the the detector housing (marked with solid blue lines). }
\label{fig:f3}
\end{figure*}

The considerations outlined above suggest that a highly detailed simulation of the detector response, from gas ionisation by the reaction products to the collection of the ionisation electrons by the anode wires, which would require interfaces to other simulation packages such as GARFIELD \cite{gar},  is not needed in order to be able to use the model to describe realistically the spatial resolution of the detector. Instead, we decided to hand-build the sensitive area of the detector by using voxels made of the Ar-CO$_2$ counting gas. The shape, size and position of the gas voxels in the counter resemble closely the readout voxels in the real detector.  Due to the particular geometry of the cathode strips, these have trapezoidal shapes. In this work we used the $G4GenericTrap$ geometrical primitive \cite{g4} to model the gas voxels and defined them as generic trapezoids described by eight vertices and a twist angle $\alpha_n$ given by $\alpha_n$  = $n\cdot\alpha$, with $n$ denoting the strip number counted from the center of the segment and $\alpha$=0.68$^\circ$ (= $\Delta \phi$, i.e., the cathode segmentation angle, see previous section). The gas voxels were arranged side-by-side to fill the space between the segment housing and the cathode plane, without overlapping, as shown in Fig. \ref{fig:f3}.  Thus, each counter of the detector segment consists of 32$\times$64 = 2048 individual gas voxels corresponding to the same number of readout channels, which leads to 2$\times$2048 = 4096 gas (readout) voxels per detector segment.  For a detection system consisting of hundreds of segments this results in a very large number of volumes to track per primary neutron.  However, by modelling the sensitive area of the detector as a collection of GEANT4 primitives rather than interfacing with GARFIELD for detailed gas phenomena calculations avoids what could become prohibitive coding and computing times, which could restrict the applicability of the model and limit the scope of the activities described in this paper. 

When the detector segment is fully assembled and placed in the detector frame, the location of each voxel is described by the spherical coordinates $r_c$, $\theta_c$ and $\phi_c$ of its centre relative to the sample position and beam axis.  The smallest gas voxels is located at the front of the detector (its centre is given by the position of the first anode wire) and $\phi_c\sim$0$^\circ$ and the largest at the back of the detector (last anode wire) and $\phi_c\sim\pm$22$^\circ$. The voxels with the coordinates ($r_c$, +$\phi_c$) and ($r_c$, -$\phi_c$) have the same size and equal but opposite sign for the twist angle  (i.e., $\pm\alpha_n$).   

During the implementation of the detector geometry in the model continuous qualitative visual inspection was carried out using the GEANT4 OpenGL viewer, checking that gas voxels do not overlap and are correctly located within the segment volume. Geometry overlap was also tested by shooting a beam of $geantinos$ (the GEANT4 debugging pseudo-particle \cite{g4}) from three orthogonal directions and perpendicular to all faces of the segment. The tracking information collected from these test runs was inspected  so as to verify which volumes were traversed by the geantinos and what is the length of each individual track.   


Rather than creating histograms and spectra directly from the simulation, the tracking information such as  the $x_i$, $y_i$ and $z_i$ positions of the interaction point and the total energy $E_i$ deposited by the charged particle in the voxel of origin were saved in a file. This output is handled with a  custom-made ROOT \cite{root} based program that stores all the position, energy deposited by the reaction product and time-of-flight information for each neutron event in a {\it root tree}, i.e., a data structure that holds many data records of the same type. This solution for storing the output data was chosen in order to profit from the versatility and capabilities of the ROOT framework for fitting, plotting as well as combining multiple data files \cite{root}. However, the output data could be easily made to have the same structure and format as the data that will be produced by the instrument in operation and could, therefore, be handled by the same visualisation/analysis software as the real data, such as MANTID \cite{mantid}. Separating the simulation part from the data analysis and representation has the benefit that the collected tracking information can be re-used to test different analysis/reduction procedures.  

In the real detector the incident neutron is localised in the voxel defined by the hardware coincidence between the wire and cathode strip signals \cite{ModzelT}.  In the GEANT4 simulation, the precise positions of the interaction  $x_i$, $y_i$ and $z_i$ and the calculated energy deposited in gas by the reaction products generated when running the code were used to establish which voxels were hit in which segment and module of the detector.  In all Jalousie variants, the wire pitch and the cathode strips and consequently, the gas voxels, have dimensions that are close to the range of the $^7$Li and $\alpha$-particles in the Ar-CO$_2$ counting gas ($\sim$10 mm \cite{SRIM}). The electron diffusion in the Ar/CO$_2$ gas mixture is far below 1 mm \cite{diffusion}, which is negligible when compared to the size of the  Jalousie voxel. Thus, a neutron that was captured by a $^{10}$B-nucleus  is expected to induce signals in the two voxels that are closest to the interaction point. Neutrons that generate tracks that end in the same voxel in which they were captured (also called interaction voxel) were assigned the coordinates of the centre of this voxel. When the tracking data indicated that the reaction product originated in one voxel but came to rest in the neighbouring one, the detected neutron was localised at the position of the centre of the voxel that has recorded the largest part of the subsequent product's kinetic energy. 

In the data reduction part of the simulation performed for the study presented in subsection \ref{subsec:res_f}, the coordinates $x_c$, $y_c$ and $z_c$ of the center of the interaction voxel were used to determine the flight-path $r_c$ given by $\sqrt{x_c^2 + y_c^2 + z_c^2}$, the angles 2$\theta_c$ and $\phi_c$ and the wavelength $\lambda_c$ for each detected neutron. The time-of-flight information was taken to be equal to the global time of the neutron track at the moment of the capture by the $^{10}$B nucleus. The drift time of the ionisation electrons\footnote[1]{The total drift time estimated with GARFIELD \cite{gar} for the ionisation electrons is about 2.5 $\mu$s for an operating voltage of 980 V and a gas mixture of 80$\%$ Ar and 20$\%$ CO$_2$ at atmospheric pressure \cite{Hen12}. The charge collection is fastest for the ionising electrons created in the region above the sense wires where the electric field has the highest values and slowest for the electrons generated in the regions with weak electric field above the field wires \cite{charpak}.} and the delays introduced by the readout electronics and the data acquisition systems were considered negligible compared to the neutron flight-time from the moderator to the detector.  The data reduction step also included the integration  over the polar angle $\phi_c$ and detection depth $z_c$ so that only the variables 2$\theta_c$, $\lambda_c$  and $d_c$ remain for each detected neutron, which is similar to the refinement of the real experimental data in MANTID \cite{mantid}.  

\subsection{Validation of the gas-voxel model through comparison with the experimental results}

For the validation of the gas-voxel model we used the results of the neutron beam tests performed with the prototype-0 of the Jalousie segment for the POWTEX instrument and reported in Refs. \cite{Hen12} and \cite{ModzelT}. In the POWTEX design the cathode is segmented into 192 strips, which are shaped such that when the segment is mounted in the holder at 80 cm from the sample, the strips pointed concentrically to the sample and delivered a constant resolution $\Delta2\theta$=0.47$^\circ$. Two wire planes, each consisting of 32 sense-field wire pairs strung perpendicularly to the cathode strips, provide signal amplification and collection. Every pair of two neighbouring signal wires are electrically connected and read out by one preamplifier channel. The electric field inside the segment and the readout scheme leads to 192$\times$16 voxels per segment counter \cite{ModzelT}. 

The measurements for testing the wire and strip resolution, corresponding to the TOF- and 2$\theta$-resolution, respectively, for the prototype-0 segment were performed at the TRIGA reactor in Mainz \cite{triga}. The reactor beam was collimated down through a set of slits with an opening of 1.6 mm, which defined a spot with a divergence of 0.51$^\circ$. This added 5.1 mm to the width of the beam on the detector located a further 400 mm downstream. A series of measurements were undertaken by moving the  segment  up-down and left-right in small steps, corresponding to scans across the wires and strips, respectively \cite{Hen12}. 

In order to be able to use the TRIGA results for the prototype-0 Jalousie segment to validate the gas-voxel model proposed in this work, the geometry of the prototype-0 segment needed to be implemented in GEANT4.  The software replica of the experimental setup also included a Gaussian beam of cold neutrons with same spatial characteristics as for the beam used in the real test. This theoretical setup was used to run a number of calculations consisting of scans along and across the wires in order to determine the spatial resolution of the segment. After each calculation the information on the number of events collected in each voxel located in the area illuminated by the beam was saved in a table. 

\begin{figure*}[ht]
\centering
\includegraphics[scale=0.4]{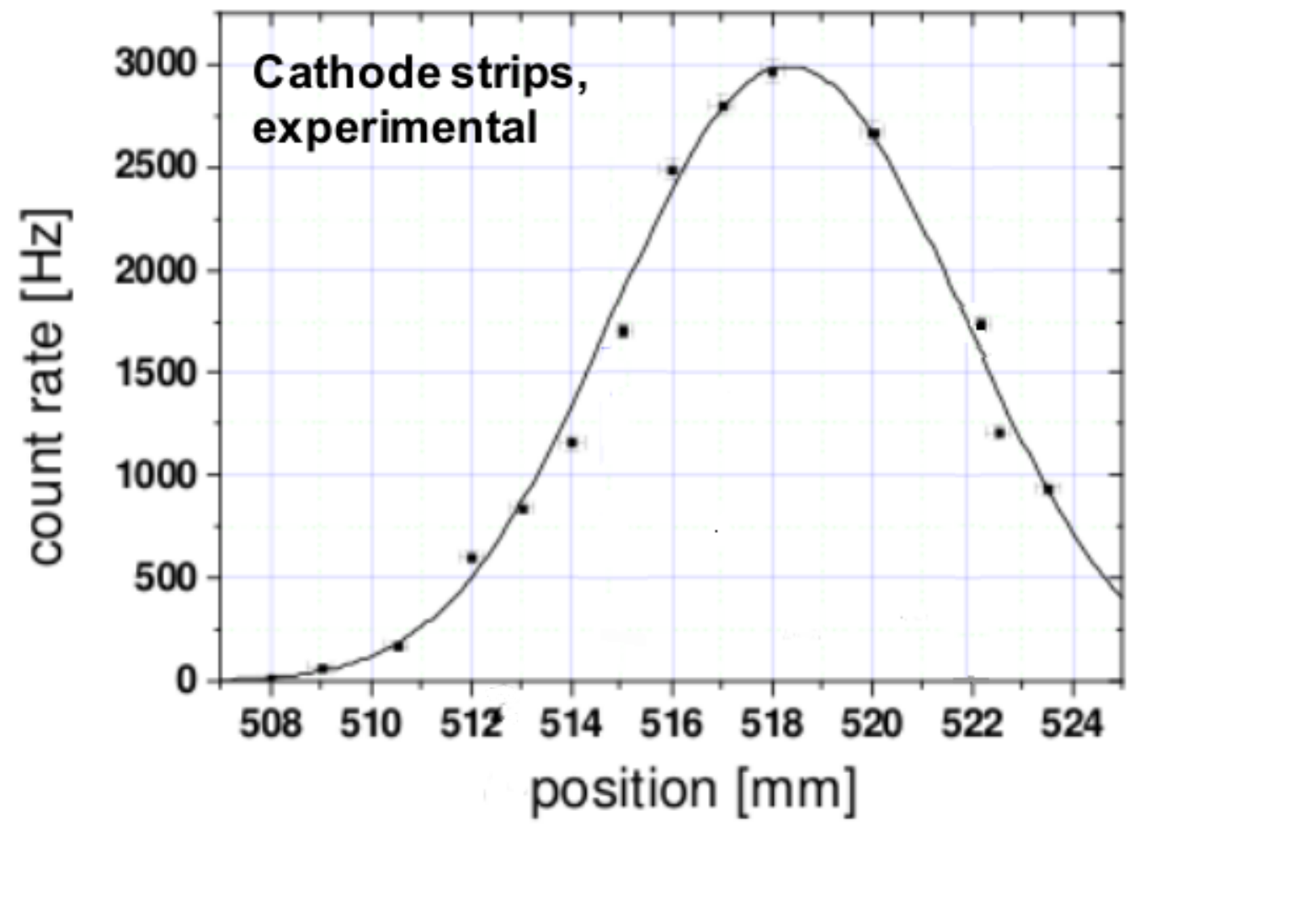}
\includegraphics[scale=0.35]{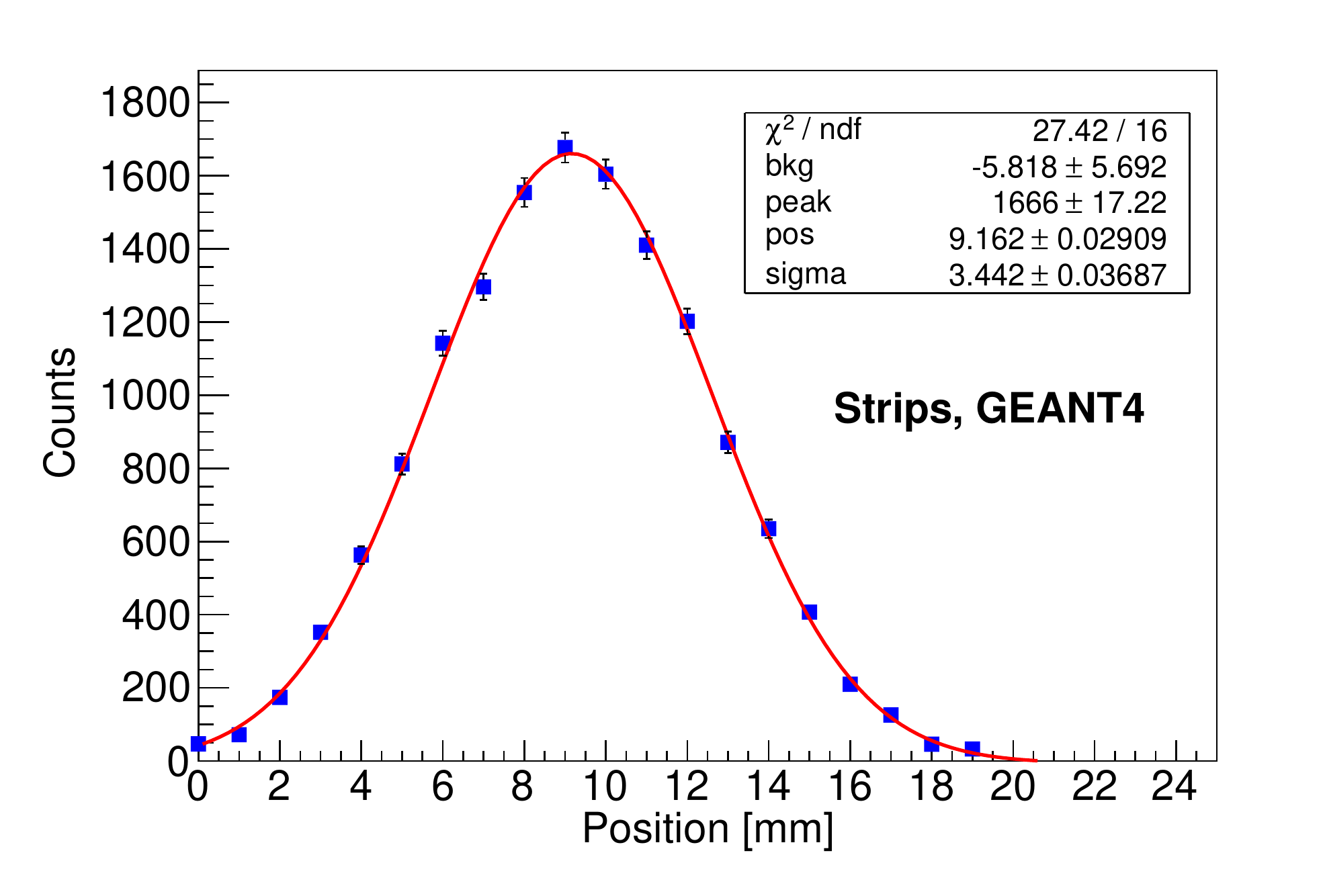}
\caption{Left: experimental intensity distribution collected during the measurements from Ref. \cite{ModzelT} in which a collimated beam was used to scan across the cathode strip (i.e., in the 2$\theta$ direction) with the width of  7.22 mm. The full line represents the Gauss fit of the experimental data (full circles), see text for details. Right: results of the GEANT4 simulation of the experimental setup used in the reactor test described in \cite{ModzelT}. The  full blue squares correspond to the number of events collected in the gas voxel with the same width as that of the experimental voxel. The result of the fit with a Gauss function is shown by the red line. The fit parameters are given on the figure. }  
\label{fig:f4}
\end{figure*}

The experimental distribution of events recorded during the reactor scan of the strip with a width of 7.22 mm is shown in the left panel of Fig. \ref{fig:f4}.  The fit of the observed distribution with a Gauss function delivered a value for the FHWM of 6.4 mm, which is larger than the physical limitation\footnote[2]{For tracks emitted orthogonal to the cathode strips, one expects, in principle, that only one strip collects the electrons liberated along the track. This corresponds to a step function distribution around that strip with a variance $\sigma$ =  strip size/$\sqrt{12}$ and FWHM =2.35$\times\sigma$  \cite{charpak2}.} of 2.35$\times$7.22 mm/$\sqrt{12}$ = 4.9 mm. The difference between the observed and calculated value is explained in Ref. \cite{ModzelT} to arise from the long track ($\sim$10 mm) of the charged particles  in the Ar-CO$_2$ counting gas. 

The right panel of Fig. \ref{fig:f4} shows  the calculated distribution of events for the voxel with the same width as the one used in the real experiment. The fit with a Gauss function of the calculated intensity distribution delivered a $\sigma$-value of 3.44(4) mm, which gives a FWHM of 8.09(9) mm. After the quadratical subtraction of the beam width of 5.1 mm, a value of 6.31(9) mm is obtained for the FWHM of the strip itself, in good agreement with the measured response. 

The calculated and measured distribution of events collected during the scans along the wires were fitted with the function given in Ref.  \cite{ModzelT}: 

\begin{align}
\label{eq:eqerf}
f(x) = \frac{A}{2}\bigg[erf\bigg(\frac{x-\bar{x_2}}{\sqrt{2}\sigma}\bigg) - erf\bigg(\frac{x-\bar{x_1}}{\sqrt{2}\sigma}\bigg)\bigg],
\end{align}

which is derived from the convolution of a Gaussian distribution with two step functions representing the slit. In the above fit function,  $erf$ is the usual Gaussian error function and the running variable $x$ is the position measured in millimetres across the wire. The fitted parameters are 
the mean positions $\bar{x_1}$ and $\bar{x_2}$, the spatial resolution of the detector $\sigma$, and the amplitude $A$ of the observed signal. 

\begin{figure*}[ht]
\centering
\includegraphics[scale=0.4]{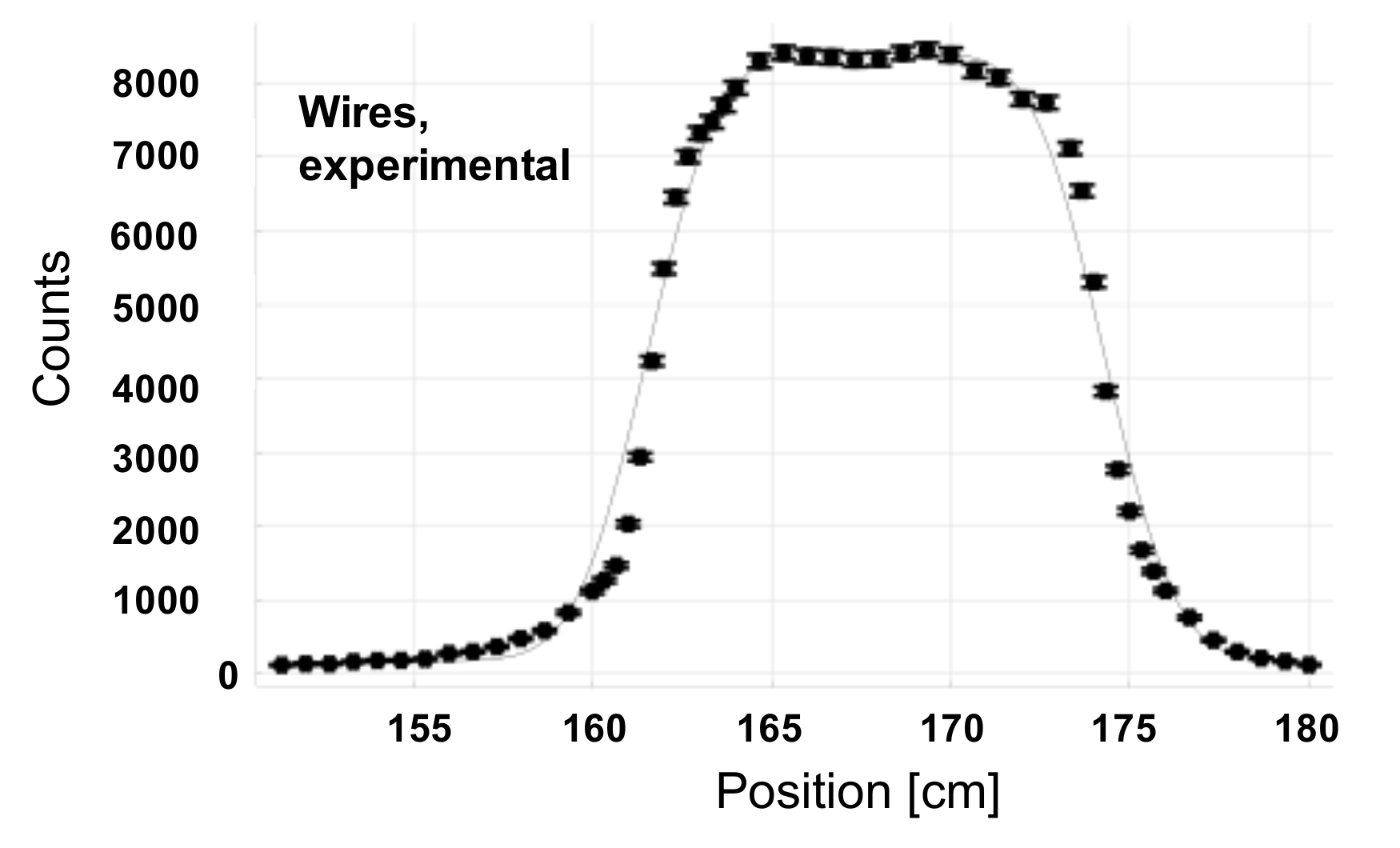}
\includegraphics[scale=0.35]{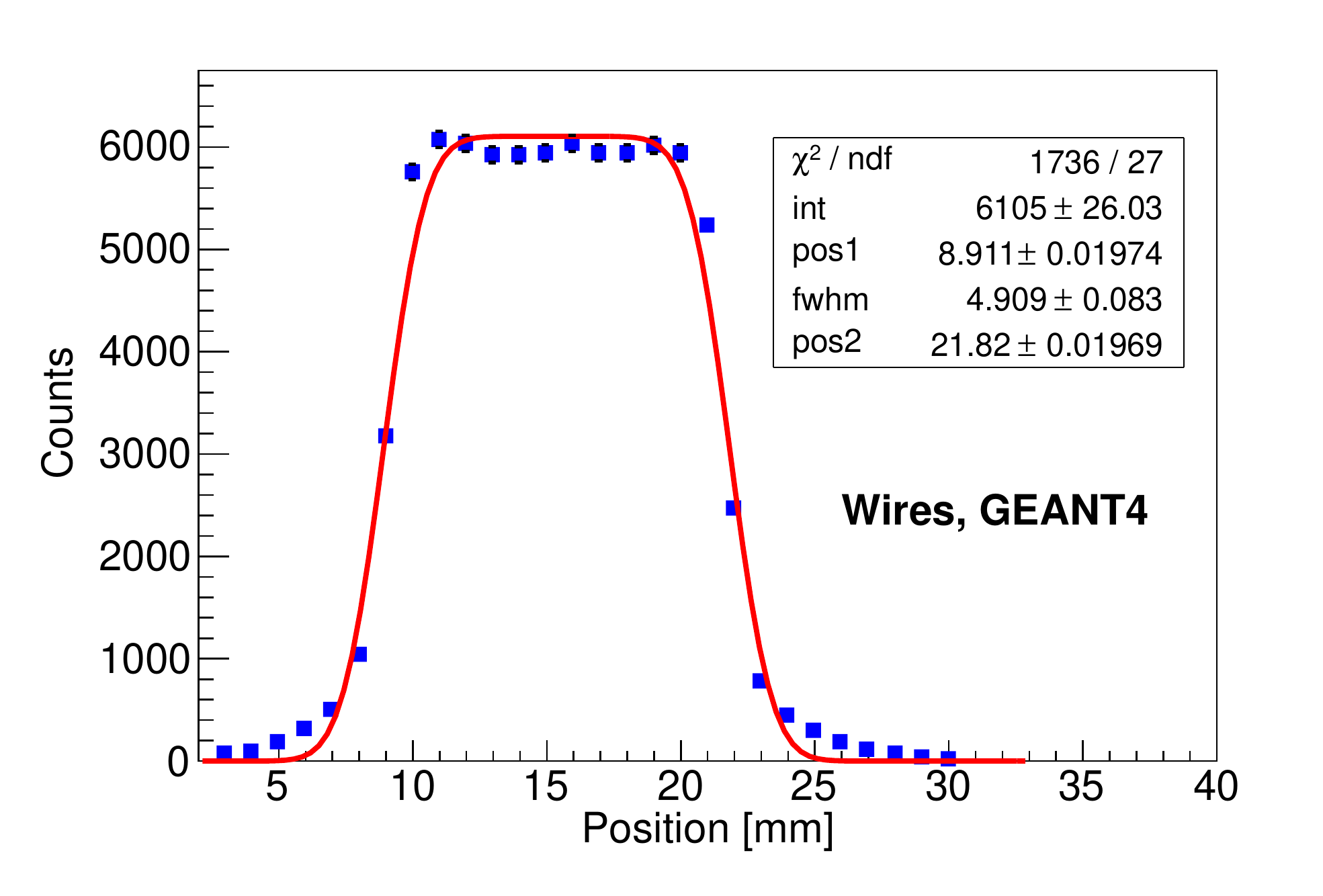}
\caption{Left: experimental intensity distribution collected during the measurements from Ref. \cite{ModzelT} in which a collimated beam was used to scan across the anode wires (i.e., along the depth of the detector, which corresponds to the TOF resolution). The experimental distribution, represented with closed square symbols, was fitted with the convolution of a Gauss distribution with two step function representing the slit, see text for details. Right: results of the GEANT4 simulation of the experimental setup and measurement of the wire resolution (full blue squares) and the results of the fit (red line). The fit parameters are given on the figure. }  
\label{fig:f5}
\end{figure*}

The experimental and calculated intensity distributions obtained from the scan across the wires and the results of the fits are shown in Fig. \ref{fig:f5}.  A value of 11.9 mm is reported in Ref. \cite{ModzelT} for the FWHM of the distribution of the signal collected by scanning across the anode wires of the POWTEX prototype segment with a wire pitch of 12.7 mm. The simulation delivers a distribution with a value for the FWHM of 12.91(3) mm given by the difference $pos2 - pos1$ of the centroids, see the inset of the plot shown in the right panel of Fig. \ref{fig:f5}. This results in a value of 12.06(3) mm (FWHM) for the wire resolution when the contribution of the Gaussian beam is subtracted. 

The good agreement between the experimental data and the results of the simulations for both the strip and wire spatial resolution presented in this subsection suggest  that the implementation of the detector geometry in the code and the simulation strategy used in the present work are correct.  Therefore, we are confident that the theoretical results presented in the next sections are reliable and can predict well the behaviour of the real detector.

\subsection{Simulation of the position resolution of the HEIMDAL-Jalousie segment}

Similar theoretical scans with collimated beam were performed with the model of the HEIMDAL segment in order to extract the FWHM for the horizontal position resolution ($\Delta2\theta$).  As in the HEIMDAL instrument the segments are arranged vertically around the sample (i.e., rotated by 90$^\circ$ compared to POWTEX \cite{Modzel}), the resolution in 2$\theta$ in this case is  determined by the wire pitch. 

\begin{figure*}[ht]
\includegraphics[scale=0.37]{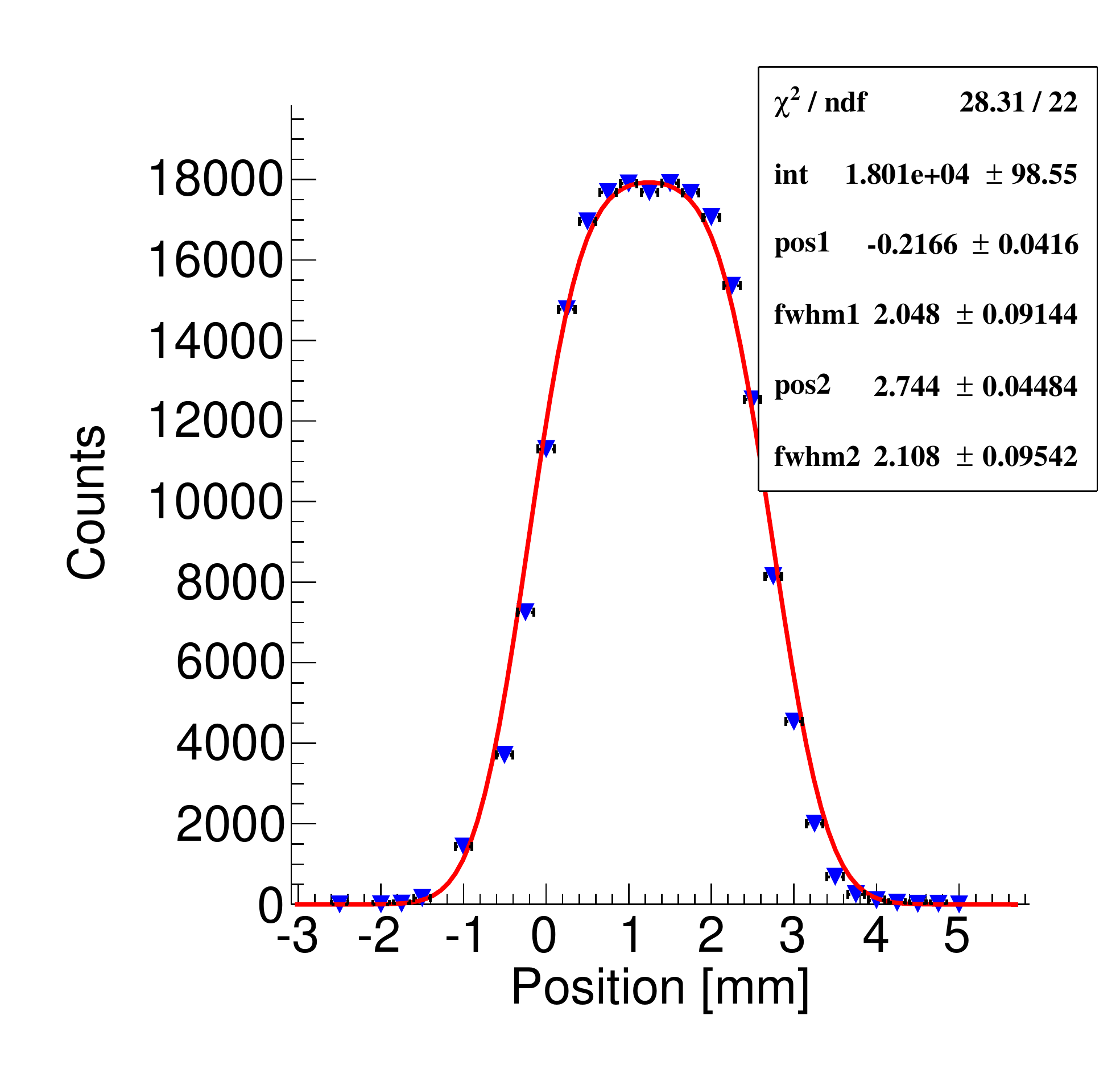}
\includegraphics[scale=0.37]{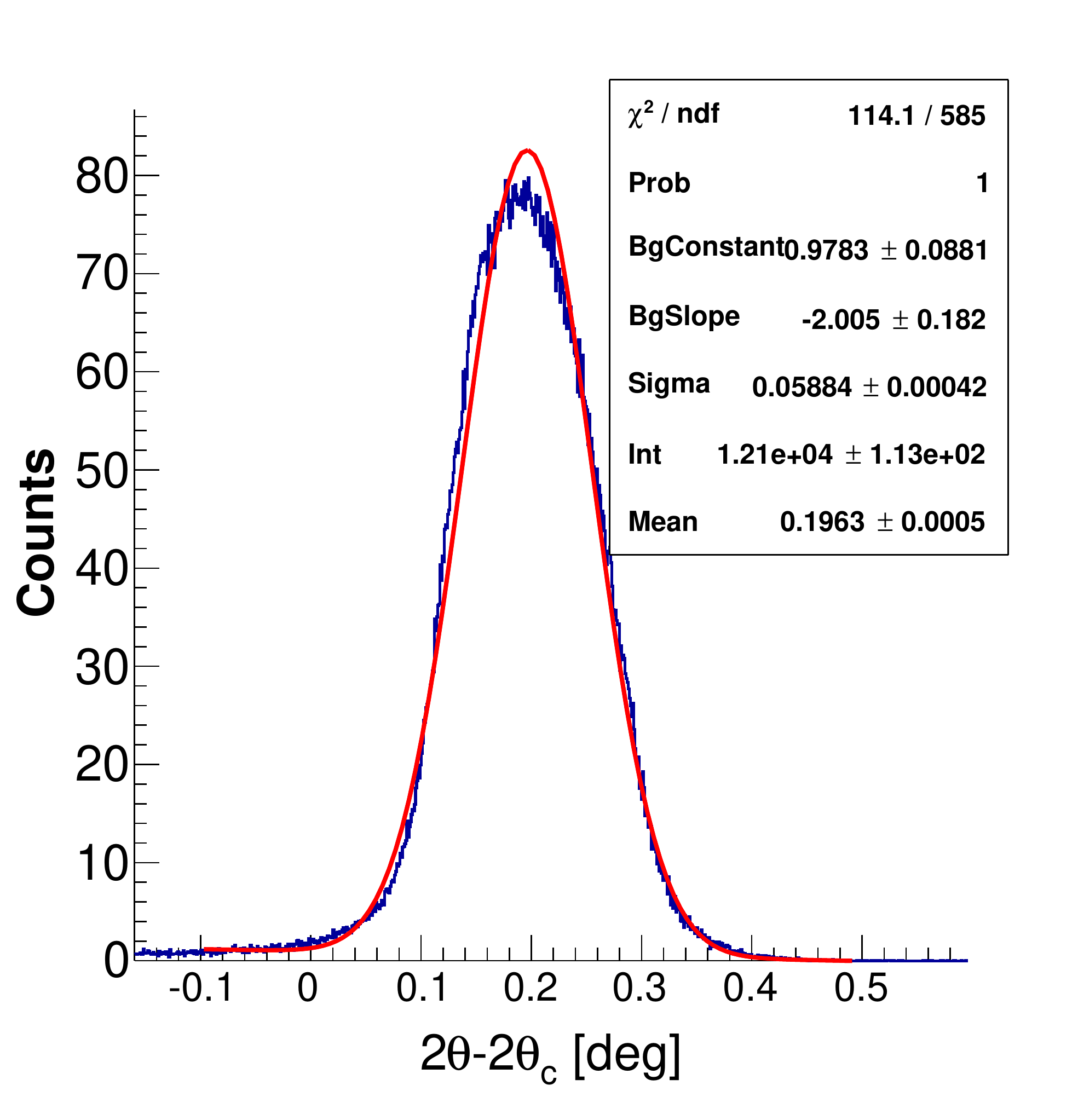}
\caption{Left: Calculated intensity distribution obtained by scanning with collimated beam across the wires of the HEIMDAL segment (blue triangles) fitted with the sum of two error functions (red line). The fit delivers a FWHM of 2.96(6) mm given by the difference between the centroids $pos1$ and $pos2$. Right: Calculated $\Delta2\theta$ for the HEIMDAL segment (blue line) fitted with a gaussian (red line).  2$\theta$ and 2$\theta_c$ represent the precise angle between the capture point of the neutron and the beam axis and the centre of the interaction voxel and the beam axis, respectively.  }  
\label{fig:f6}
\end{figure*}

The results of the simulation for the $\Delta$2$\theta$-resolution for the HEIMDAL segment baseline design are shown in Fig. \ref{fig:f6}. Same as in the validation analysis presented in the previous subsection, the intensity distribution calculated by scanning with collimated beam across the width of the segment and shown in the left panel of the figure was fitted with the function given by Eq. \ref{eq:eqerf}. In this calculation, the profile of the incident neutron beam was assumed to be a Gaussian with both the horizontal and vertical width of 0.47(1) mm. This added an additional contribution of 1.11(2) mm to the FWHM of the distribution of intensities collected by the voxels in the area of the detector illuminated by the beam. The  quadratic subtraction of the beam contribution from the difference $pos2$ - $pos1$ gives a value 2.74(3) mm for the position resolution (FWHM) in the horizontal (scattering) plane. The right panel of the Fig. \ref{fig:f6}  shows the residual 2$\theta$ - 2$\theta_c$, where 2$\theta$ and 2$\theta_c$ are the angles between the capture point of the neutron and the centre of the voxel that collects the signal with respect to the beam axis, respectively. The 2$\theta$-angle was determined by using the Monte-Carlo coordinates $x$, $y$, $z$ of the interaction (capture) point of the incident neutron, while 2$\theta_c$ corresponds to the centre of the interaction voxel. A value of 0.138$^\circ$  for the FWHM was extracted from the Gauss fit of the calculated residual distribution. This value agrees well with the angular resolution required by the HEIMDAL instrument team in order to be able to perform structural studies \cite{heimdal,Holm}. 

The good agreement between the requirements set by the instrument team for the detector performance and the simulation results presented in this section validates the baseline design for the HEIMDAL detector and it could be considered the first good indication that the selected detection solution is able to carry out the scientific goals. 

 \subsection{The resolution function of a powder diffractometer}\label{subsec:res_f}

The success of a powder diffractometer relies in its ability to provide data in which the Bragg reflections have good statistics, but at the same time are well-separated over a wide $d$-range. This requires a great deal of optimisation of the instrument design with the help of computer models such as McStas or VITESS \cite{mcstas,vitess,restrax}.  One figure-of-merit that is carefully looked at during the optimisation process is the instrument resolution function, denoted as $\Delta d/d$, where $\Delta d$ is the width of a Bragg reflection for a given $d$-value. Higher resolution (i.e., low $\Delta d/d$ values) gives sharper peaks, which means that sample information, such as strain and dislocation density, can be more accurately determined. The broadening of the Bragg reflection is due to the convolution of the intrinsic peak width from the sample itself with an instrumental contribution called instrument resolution function or response function \cite{lpf}.  Knowledge of the instrumental contributions to the peak broadening is important in order to evaluate the absolute resolving power of the diffractometer. 

Analytical approximations to estimate the response function exists for both monochromator and TOF diffraction instruments \cite{pdf}. The response function for the latter is found by using error propagation of Bragg's low for TOF, $t = 2\alpha\cdot L\cdot d\cdot sin\theta$:  

\begin{align}
\label{eq:eq_res}
\frac{\Delta d}{d} = \bigg[\bigg(\frac{\Delta L}{L}\bigg)^2 + \bigg(\frac{\Delta t}{t}\bigg)^2 + \bigg(\frac{1}{2}\frac{\Delta2\theta}{\tan\theta}\bigg)^2\bigg]^{1/2}.
\end{align}
where $\alpha=h/m_n=0.2528$ ms/\AA$\cdot$m, $\Delta L$ is the uncertainty  in the distance traversed by the neutrons between the moderator and detector, $\Delta t$ is the uncertainty in the corresponding time-of-flight $t$ for the distance $L$, and $\Delta 2\theta$ is the uncertainty in the scattering angle 2$\theta$. The flight path uncertainty $\Delta L$ arises from the finite size of the moderator, sample and detector while $\Delta t$ depends on  the neutron pulse length and the time resolution of the detector. The angular term $\Delta2\theta$ in Eq. \ref{eq:eq_res} contains contributions from both the neutron optics, through the slit opening, and the detection element (detector pixel) as viewed from the sample \cite{Holm}. 




The optimisation of the design of a powder diffraction instrument is usually performed by selecting the detectors near 2$\theta$ = 90$^\circ$ scattering angle and balancing the individual contributions to the resolution function such that each of the three terms in Eq. \ref{eq:eq_res} contributes with 30$\%$ to the total resolution function. This optimisation procedure is known as {\it resolution matching} \cite{Holm} and has a direct impact on the range and flexibility of the instrument.  The optimal sample size is directly tied to how large the beam divergence can become and what is the finest detector spatial resolution achievable with the current detector technologies.  Thus, the geometry (thickness, pixel size) and timing performance of the detector must be carefully considered when selecting the detector technology for a powder diffraction instrument. Scintillator-based detectors use  thin neutron converters such as ZnS:$^6$LiF(Ag) screens coupled to the light-detection device by optical fibres with 1-2 mm diameter \cite{gem}. However, using this technology to determine the position of the detected neutron with an accuracy of a few mm  is still challenging when several steradians  of coverage are required, and the rate capability is also restricted by the long afterglow of the scintillator \cite{scint}. Pressurised $^3$He-tubes have good timing resolution, but the horizontal pixel size is limited by the tube diameter to 7-8 mm \cite{wish,wish2}. Thus, for the state-of-the-art detector technologies employed at existing similar instruments operational at the available neutron sources, the detector pixel size is regarded as the main limiting factor in the performance of a diffraction instrument used to investigate small samples in high-resolution mode. 

The resolution calculations and optimisations that constituted the basis for extracting the detector requirements for the HEIMDAL instrument are presented in Ref. \cite{Holm}.  For this particular instrument, the first term of Eq. \ref{eq:eq_res} is negligible as this will be one of the longest powder diffractometers ever built ($\sim$157 m).  The contribution of the second term of the equation to the resolution function depends on the opening time of  the pulse-shaping choppers, which will be 123 $\mu$s in the high-resolution mode of operation \cite{Holm}.  In addition to the width of the neutron pulse one should also consider the time uncertainty of $\sim$8 $\mu$s due to the flight time of the 1.8 \AA~neutrons through the Jalousie voxel, see subsection \ref{subsec:jal_h}, which results in a $\sim$6.5$\%$ additional contribution to $\Delta t$. 

Experimentally, the instrumental resolution of a neutron diffractometer is usually determined in diffraction measurements with standard calibration samples such as  Na$_2$Ca$_3$Al$_2$F$_{14}$, which is a powder with very well-known crystallographic structure that exhibits a number of resolvable peaks with low intrinsic line widths over a wide $d$-range \cite{Langford}. The experimental diffractogram is fitted with the theoretical pattern calculated for a perfect powder convoluted with a peak shape depending on the characteristics of the beam and instrumental parameters. The knowledge of the experimental resolution function is also important as a check for the alignment of the instrument or to benchmark the performance  of different neutron diffractometers. For a TOF diffractometer, the resolution $\Delta d/d$ decreases with increasing angle $\theta$ (see Eq. \ref{eq:eq_res}), but it depends weakly on $d$ for a given scattering angle. 

It is worth mentioning that some TOF diffraction instruments such as GEM \cite{gem1} and POLARIS \cite{polaris} have their detectors arranged in the so-called \emph{resolution focussed} geometry, which consists of several discrete detector banks mounted around the sample under investigation. Each bank covers a specific scattering angle, which leads to an approximately constant term 1/$\tan\theta\Delta2\theta$ in Eq. \ref{eq:eq_res} and therefore, approximately the same resolution for all detector elements within a given bank. For this reason, the data sets collected from the different banks are treated separately when applying the normalisation procedure \cite{gem1,polaris}. In contrast, by arranging their detectors on a cylindrical locus around the sample, the HEIMDAL, DREAM and POWTEX instrument teams aim to benefit from the largely varying resolution function with both $\theta$ and $\lambda$ to perform a novel 2D refinement method, as explained in detail in Refs. \cite{2drefine1,2drefine2}.   

\subsection{Study of the contribution of the detector to the instrumental resolution function }

We can use the GEANT4 model of the HEIMDAL-Jalousie detector to estimate the contribution of the baseline design to the instrument resolution function (Eq. \ref{eq:eq_res}) compared to a state-of-the-art detector technology that has  been in operation for several years and proved to be reliable and deliver high quality diffraction data.  The reference detector chosen for this study is that employed at the WISH instrument \cite{wish}, one of the world leading powder diffractometers operational at the ISIS spallation source \cite{isis}. Please note that here we do not intend to compare the performance of the two diffraction instruments, but two detector technologies for a specific instrumental arrangement. 

The WISH detector consists of 1520 pressurised $^3$He position-sensitive tubes with 8 mm diameter and a height of 1 m mounted in two semi-cylindrical annuli around the sample. The height of a tube is electronically divided into 128 channels, each having a nominal resolution of 8 mm, which matches the horizontal spatial resolution given by the tube diameter \cite{wish}. 

The GEANT4 model of the WISH detector was built in this work by following the recipe described in section \ref{g4_section}. Same as for the HEIMDAL detector, the position sensitivity along the tube was modelled by stacking 128 cylindrical gas voxels with a diameter of 8 mm and a height of 8 mm. The WISH detector model was used to perform two sets of calculations. In the first calculation, the tubes were arranged in cylindrical geometry around the sample at 220 cm distance, the length of the secondary flight-path of the real WISH instrument \cite{wish,wish2}. In the second calculation, the tubes were placed at  80 cm from the sample position, same as the radius of the HEIMDAL detector cylinder.

\begin{figure*}[ht]
\centering
\includegraphics[scale=1.05]{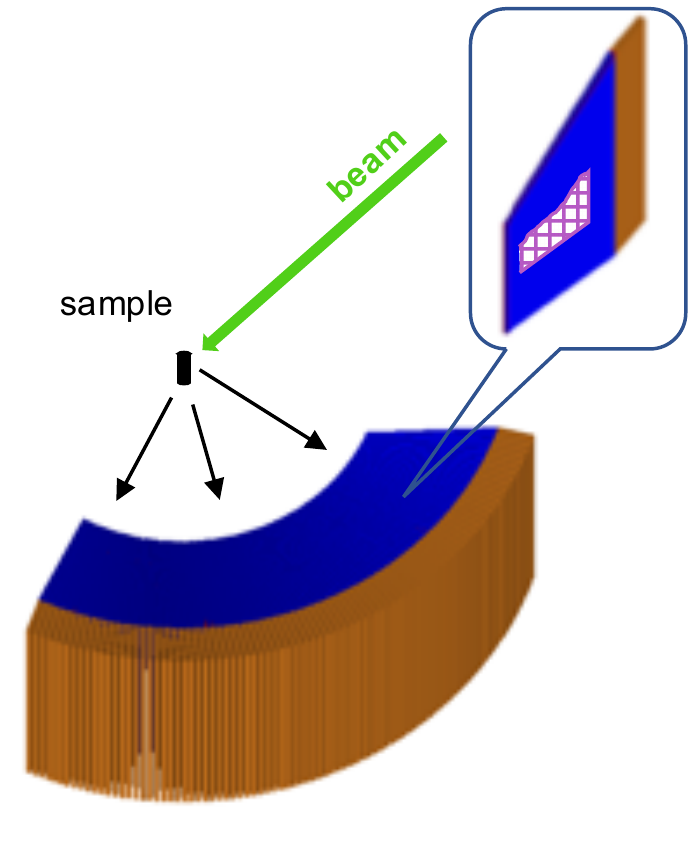}\\
\includegraphics[scale=0.75]{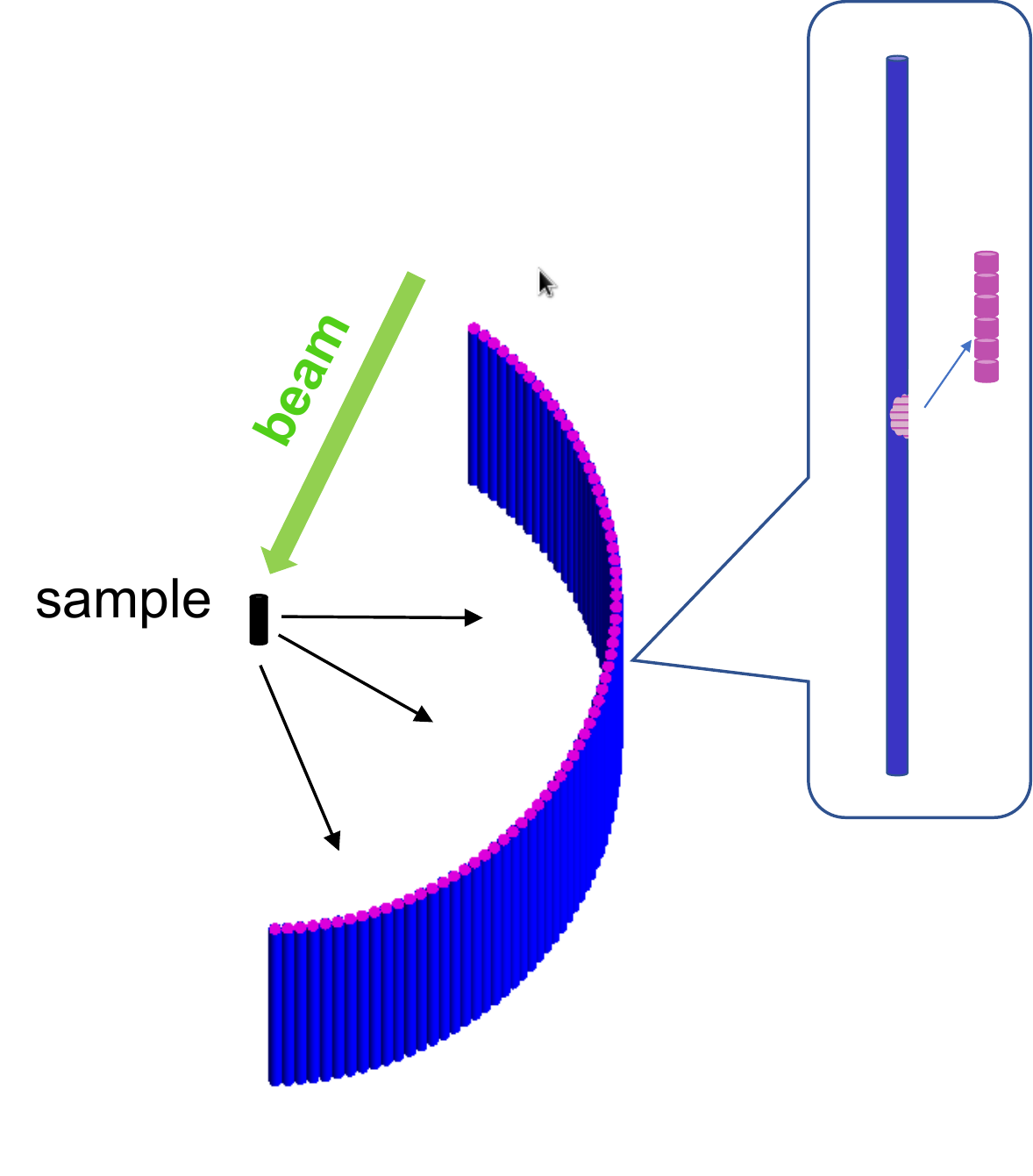}
\caption{Top: Visualisation of the geometric configuration for the HEIMDAL-Jalousie sample area used to simulate the NAC diffraction spectrum discussed in this section. Bottom: same, but for the WISH detector. The insets show a  graphical representation of the implementation in the code of the position sensitivity of the Jalousie segment and $^3$He tube by means of trapezoidal and cylindrical gas voxels, respectively. }  
\label{fig:f10}
\end{figure*}

The two detector geometries, shown schematically in Fig. \ref{fig:f10}, were used to simulate the diffraction pattern of the fluoride Na$_2$Ca$_3$Al$_2$F$_{14}$, denoted NAC hereafter, which is one of the reference samples mostly used for the commissioning and calibration of real powder diffraction instruments \cite{pdf}. The incident neutron spectrum used in the GEANT4 calculations was generated with the help of the VITESS computer package \cite{vitess}. This code uses the Monte-Carlo ray-tracing technique to simulate neutron scattering instruments. An  initial beam distribution ranging from 0.1 - 20 \AA~was generated with the help of the moderator file $IsisTS2broad.mod$ available for download with the VITESS simulation package \cite{vitess}. This component produces a time-of-flight spectrum that has the characteristics of the ISIS TS2 solid-methane moderator \cite{ch4}. The initial neutron beam was subsequently shaped and transported to the NAC sample with the help of an instrument assembled from virtual components representing choppers, slits and neutron guides. The length of this virtual instrumental, measured from the moderator up to the sample, was 41 m, which is the same as the length of the real WISH instrument at ISIS \cite{wish}, but significantly shorter than HEIMDAL. The NAC sample used in the VITESS calculation was a cylinder with a radius of 5 mm and a height of 40 mm. This size for the sample was chosen in order to ensure good statistics for a reasonable computing time. The information on the energy, position and momentum ($x$, $y$ and $z$-components) of the neutrons scattered by the powder sample were stored in a text file, which was used as input information for the GEANT4 primary generator file \cite{g4,g4a,g4b}. In the subsequent GEANT4 simulation of the diffraction patterns, the VITESS trajectories were used as primary particles originating from the centre of the detector cylinder, which coincides with the sample position in the VITESS calculation. The GEANT4 primary particles are initiated at the time $t_i$ corresponding to the time-of-flight recorded at the sample position in the VITESS part of the calculation. Thus, the values for the $d$-spacing for each detected neutron were determined from the wavelength information calculated by using the time-of-flight information given by the total time delay between neutron production by the VITESS moderator component and the time of capture by the Boron converter inside the detector. In case of neutron scattered inside the detector, the total time delay was  larger, resulting in a reconstructed energy that was lower than the primary neutron energy. These kind of neutron events contributed to the background signal. Thus, the total flight path of the scattered neutron used in the determination of the wavelength was given by the distance between the VITESS moderator and the centre of the GEANT4 interaction voxel. 

\begin{figure*}[ht]
\centering
\includegraphics[scale=0.8]{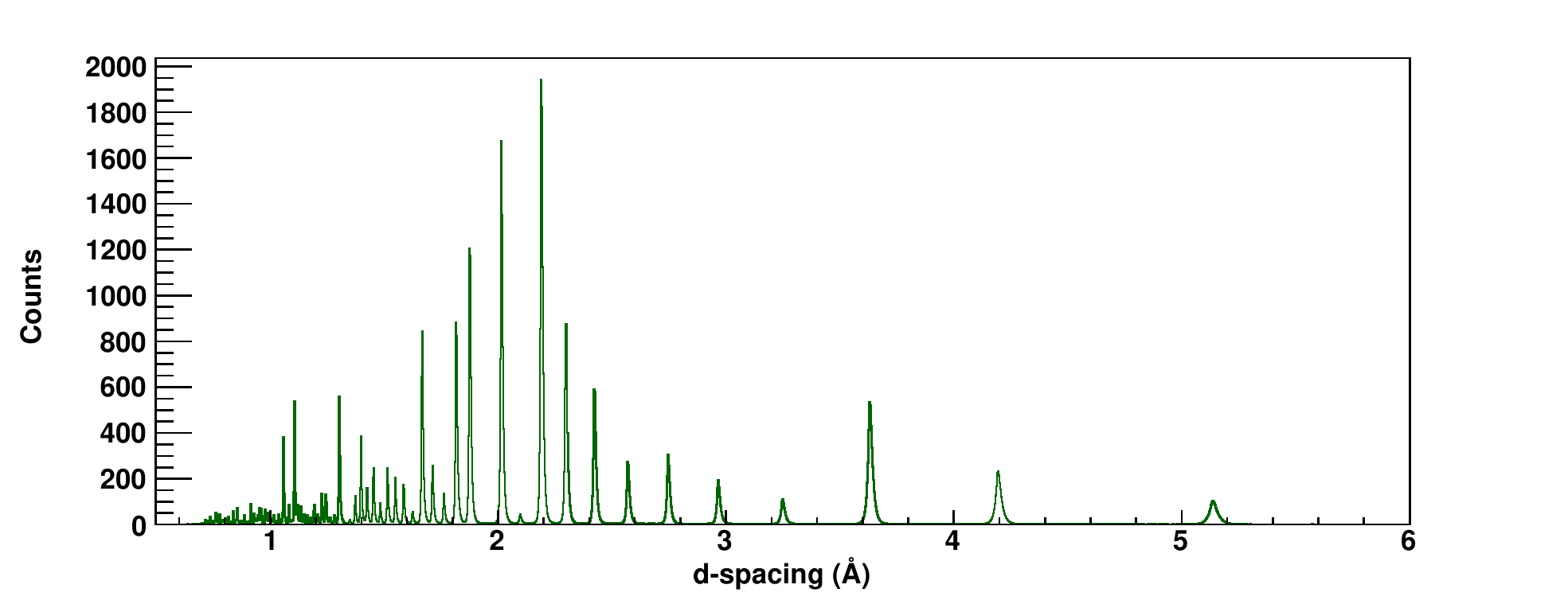}\\
\includegraphics[width=0.94\linewidth]{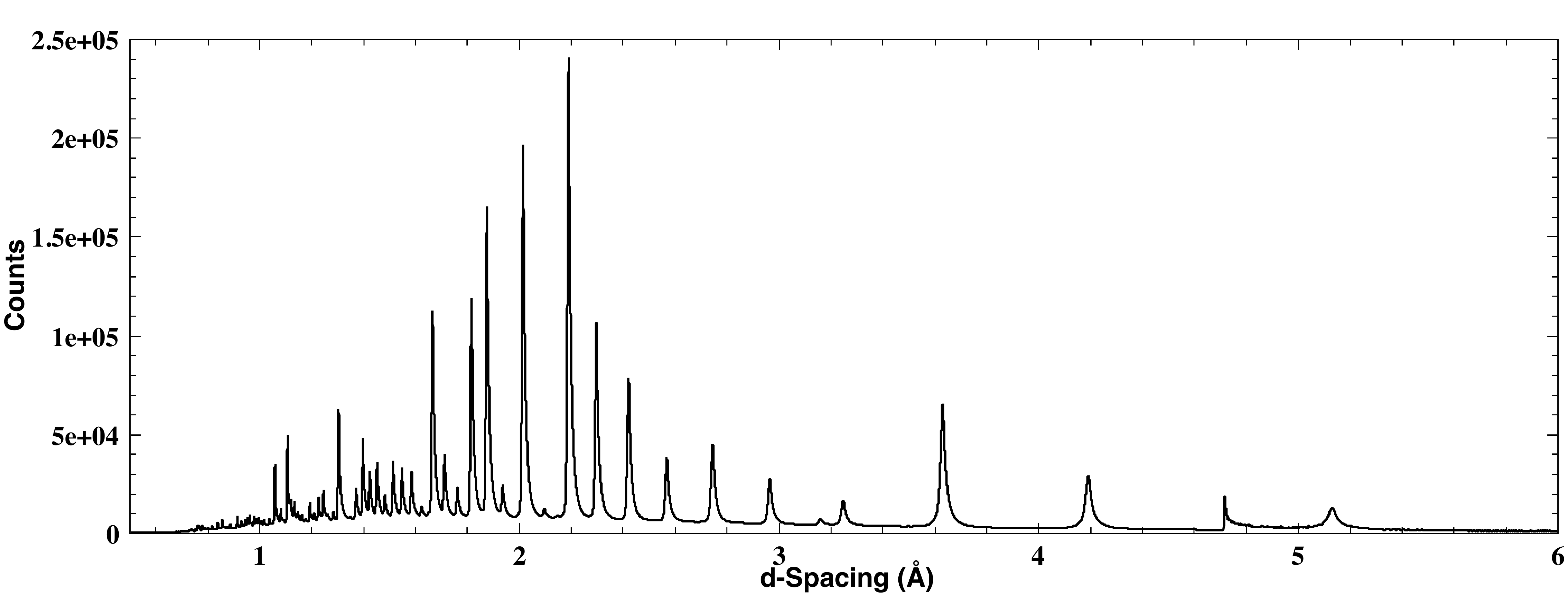}
\caption{Top: NAC diffraction pattern calculated with the GEANT4 detector model for the WISH-like detector placed at 220 cm from the sample. The neutron trajectories used as input in the GEANT4 Particle Generator were generated with the VITESS virtual model of the solid-methane (cold) moderator on TS2 and a diffraction instrument bearing similarities with the WISH diffractometer. Bottom: experimental diffraction pattern collected at the real WISH instrument.   }  
\label{fig:f11a}
\end{figure*}

The diffraction patterns for the NAC powder sample calculated with the WISH-like detector model is shown on top panel of Fig. \ref{fig:f11a} as an 1D-spectrum representing the number of counts versus $d$-spacing. For the sake of illustrating the quality of the simulation results, which is an indicator of the accuracy of the detector models used to generate the spectra, the NAC spectrum collected with the real WISH instrument is shown on the bottom panel of the same figure. The experimental spectrum was obtained by summing up the raw diffraction signals (i.e., with no corrections applied to the data, such as detector dead-time, attenuation, multiple scattering, etc.) collected from all of the 1520 $^3$He-filled tubes of the WISH detector \cite{wish,wish2}. As seen in Figure \ref{fig:f11a}, the weights and positions of the calculated Bragg peaks agree very well with those observed experimentally. The background under the peaks in the simulated spectra arises from the scattering of the neutrons in the detector materials only, which is not well treated in the default GEANT4 \cite{g4c}.  Therefore, it is much smaller than the background observed in the real WISH detector, which also contains events from other sources, such as cosmic generated neutrons and scattering in the beamline components located near the beam path. The feature present around 4.7 \AA~in the experimental spectrum is an artefact caused by the sharp cut-off at 10 \AA~ in the incident neutron spectrum available during the NAC measurement. This limits the observable $d$-range to $\sim$5 \AA~ in the most  backward detectors (137$^\circ\le$2$\theta\le$167$^\circ$). 

\begin{figure*}[ht]
\centering
\includegraphics[scale=0.8]{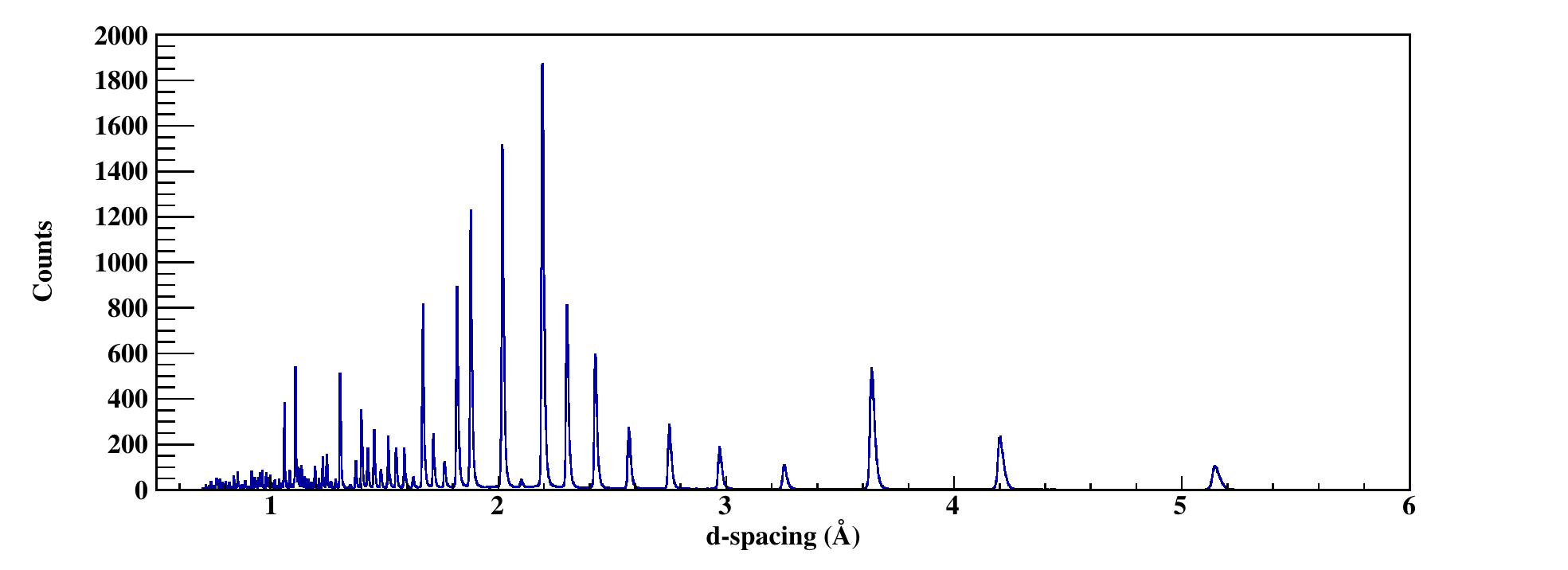}\\
\includegraphics[scale=0.8]{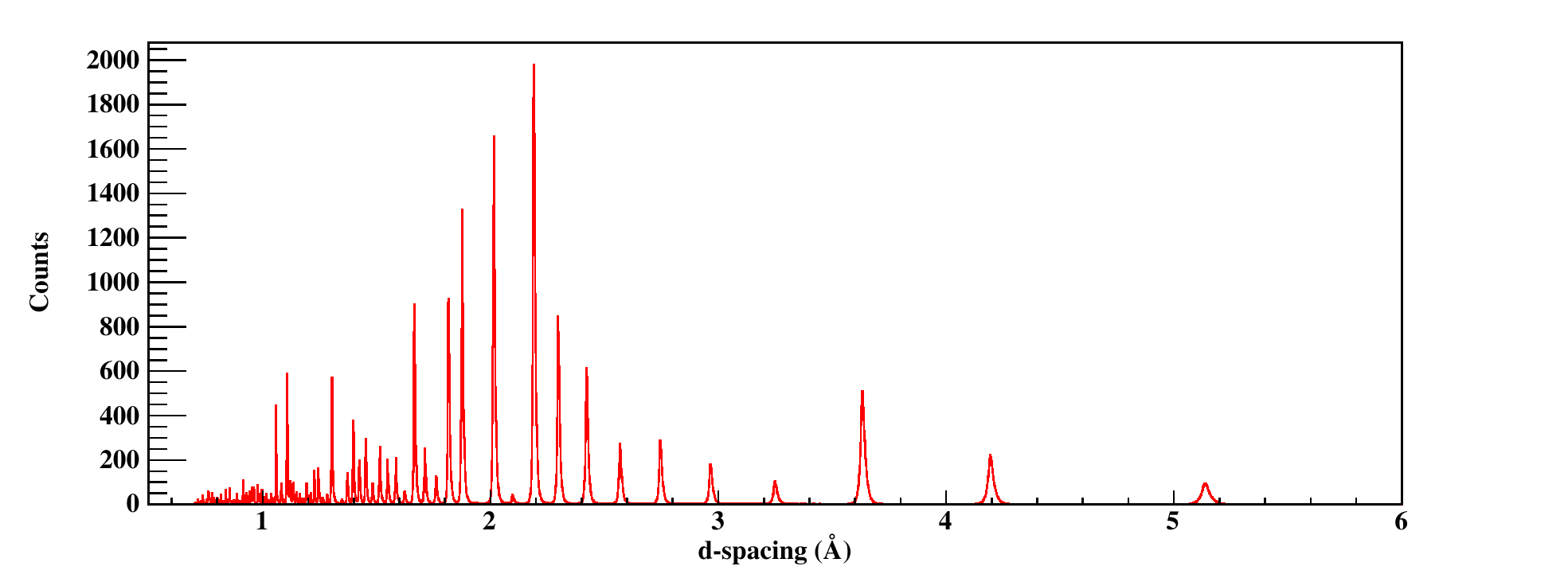}\\
\caption{Top: NAC diffraction patterns calculated with the GEANT4 detector model of the HEIMDAL-Jalousie detector. Bottom: NAC spectrum calculated with the  WISH-like detector model. Both spectra were calculated with the same incident neutron distribution (see text for details) and for a secondary flight-path of 80 cm. Please note that the NAC spectrum that will collected with the real HEIMDAL instrument will look different than that illustrated in the top panel of this figure as this instrument is designed to use the thermal part of the ESS pulse, which will give a different contribution to the weights of the characteristic Bragg reflexions.}  
\label{fig:f11b}
\end{figure*}

Figure \ref{fig:f11b} compares the diffraction patterns for the NAC powder sample calculated with the HEIMDAL-Jalousie (top panel)  and WISH-like (bottom panel) GEANT4 detector models. Both spectra were simulated for a distance from the sample to detector of 80 cm.  
Since the input neutron trajectories in the both GEANT4 simulations were generated with the same instrument (moderator) model and for the same secondary flight-path, the weights of the Bragg peaks are the same in both spectra. However, the difference in the voxel geometry (size) is expected to influence the width of the diffraction peaks, and this aspect was further investigated as described below.

\begin{figure*}[ht]
\centering
\includegraphics[scale=0.55]{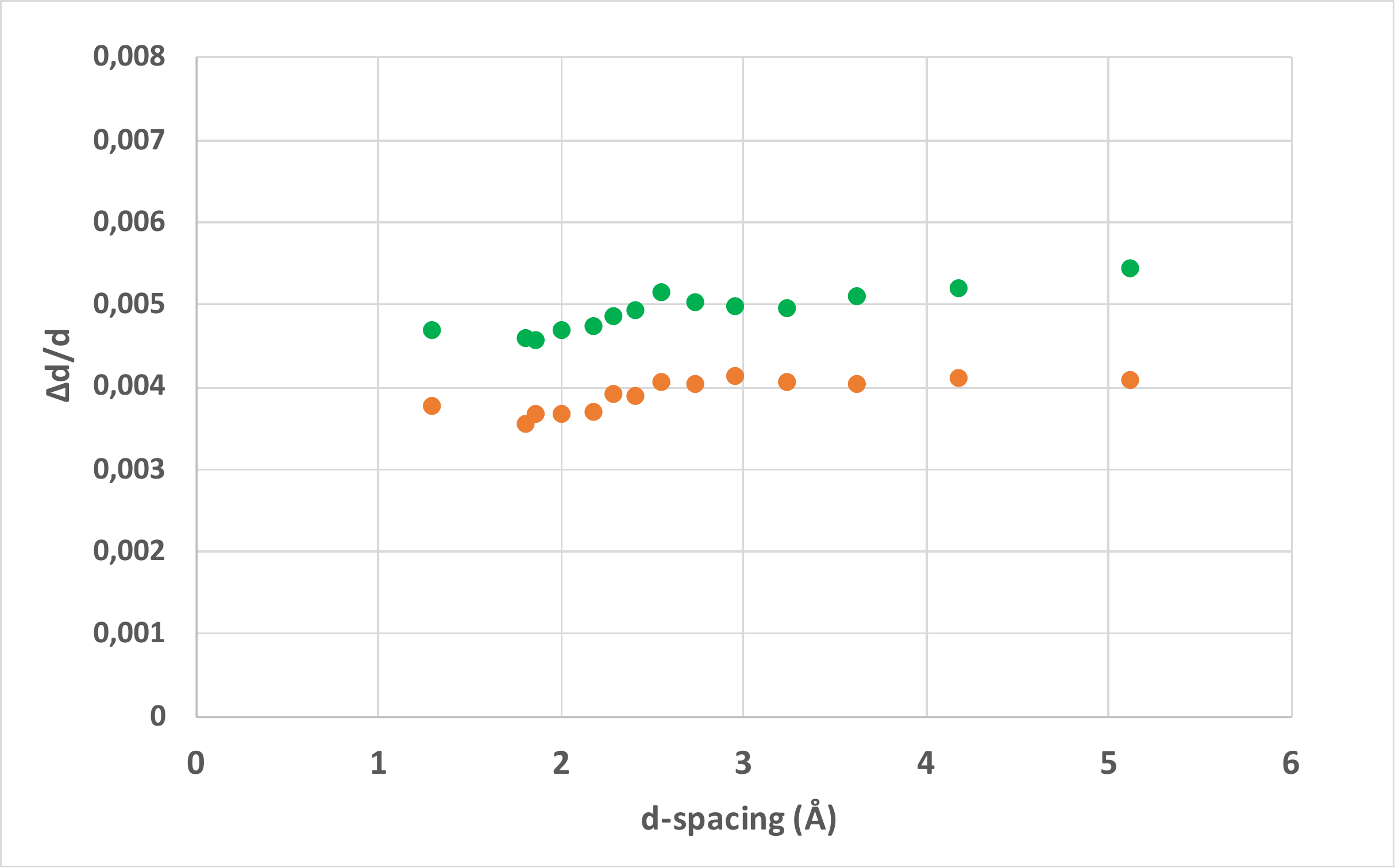}
\caption{Dependence of the $\Delta d/d$-ratios with the $d$-spacing. The ratios were extracted from the individual fit of the Bragg reflections of the NAC sample calculated with the GEANT4 models of the HEIMDAL-Jalousie (orange filled circles) and WISH-like detectors (green filled circles) located at 80 cm from the sample position. Only the data from the detectors covering scattering angles between 85$^\circ$ and 95$^\circ$ was used in the fit.}  
\label{fig:f12}
\end{figure*}

The theoretical diffractograms obtained from the data collected in the detectors located at 85$^\circ\le2\theta\le$95$^\circ$ with respect to the beam axis were used to extract the $\Delta d/d$-ratios for the most intense Bragg reflections of the NAC sample. The peaks were fitted with the function obtained by convoluting a Gauss distribution with an exponential tail that accounts for the decay characteristic of the pulse generated by the VITESS moderator module. A linear function for the underling background was also assumed. The nominator $\Delta d$ corresponds to the values for the FWHM returned by the fit of the peaks. The dependence of the calculated $\Delta d/d$-ratios with the $d$-spacing is shown in Fig. \ref{fig:f12}. As expected for a time-of-flight virtual instrument, the ratios vary only slightly with the lattice distance $d$. The small increase observed for values above 2.5 \AA~could be due to the increase of the width of the moderator pulse  with increasing wavelength \cite{ch4}. The absolute values of the $\Delta d/d$-ratios determined in this analysis are meaningless. The virtual instrument used to generate the diffraction trajectories might resemble the WISH instrument operational at the ISIS source, but the model is not accurate enough to allow for direct comparison with real data, which is also out of the scope of this work. Only the relative difference between the two data sets is relevant to our discussion as it is directly related to the instrumental contribution from the detectors. Such a comparison between the two simulated resolution functions is possible because the same instrumental setup was used to generate the primary neutron trajectories for the GEANT4 simulation and the data was analysed in exactly the same manner. The smaller ratios obtained from the simulation with the HEIMDAL-Jalousie detector model indicate sharper Bragg peaks and therefore, better instrumental resolution when using this detector technology to collect the diffraction signal. Please note that this does not imply that the HEIMDAL instrument is expected to outperform WISH. In the real WISH instrument the $^3$He detectors are located at 220 cm from the sample \cite{wish}, therefore the pixel size as viewed from the sample is smaller than in the setup used in our calculation.


\section{Conclusions} 

The scientific goals of ESS place demanding requirements on the performance of the detection systems for most of the planned instruments.  One of the several new instruments approved for construction in the first tranche is the thermal powder diffractometer HEIMDAL. The theoretical study of the performance of the Jalousie detector technology has been performed to support the design effort and engineering choices for the HEIMDAL powder diffraction detector. A detailed software replica of the Jalousie detector was built in GEANT4 by using dimensions according to the available technical drawings provided by the instrument team. In order to avoid time-consuming simulations, but still be able to study the performance of the detector in terms of spatial resolution, a gas-voxel model that matches the readout segmentation foreseen for the real system was hand-coded. The simulated detector response was made available in a simple text format, allowing further processing and event reconstruction with performant analysis software. The simulation results  have been benchmarked to reproduce spatial resolution data obtained with the Jalousie-module designed and built to fulfil the requirements of the future POWTEX diffraction instrument under construction at FRM2. Despite the simplifications made when simulating the detector operation, there is an excellent agreement between the calculated and measured intensity distributions obtained in the investigation of the position sensitivity of the POWTEX-Jalousie detector.  This provides confidence in the realism of the system description included in the GEANT4 model and demonstrates that highly reliable results can be obtained in the simulation of the Jalousie detectors for ESS. The current work also presents extrapolations for the performance of the Jalousie baseline detector for HEIMDAL in terms of contribution to the resolution function. 

The simulation results presented here indicate that the proposed baseline design for the HEIMDAL-Jalousie powder diffraction detector is close to the optimal one and give confidence that the performance required by the instrument team will be achieved. We assume that the actual detector design will not deviate considerably from the design used in this work to investigate its performance with the help of the GEANT4 simulation package, but if it does, the detector model can be easily modified to account for all changes. Also, future prospects include the implementation in the current detector model of other components relevant to the instrument end station such as the sample holder, sample environments and the experimental cave, in order to study their contribution to the background signal recorded in the detector. Obviously, our goal is to be able  as soon as possible to overlay data from the real detector onto the simulated signal events. 

\acknowledgments

This work was partially funded by EU Horizon2020 framework, BrightnESS project 676548.

\end{document}